\documentclass[english,,conference]{IEEEtran}

\newcommand{\subparagraph}{}

\usepackage{titlesec}
\titleformat*{\subsection}{\sc}
\titlespacing*{\paragraph}{0pt}{-0.5ex}{-3ex}

\usepackage{graphicx}
\usepackage{subcaption}
\usepackage{listings}
\lstdefinelanguage{JavaScript}{
  basicstyle=\ttfamily\small,
  morekeywords=[1]{break, continue, delete, else, for, function, if, in,
    new, return, this, typeof, var, void, while, with},
  morekeywords=[2]{false, null, true, boolean, number, undefined,
    Array, Boolean, Date, Math, Number, String, Object},
  morekeywords=[3]{eval, parseInt, parseFloat, escape, unescape},
  sensitive,
  morecomment=[s]{/*}{*/},
  morecomment=[l]//,
  morecomment=[s]{/**}{*/}, 
  morestring=[b]',
  morestring=[b]"
}[keywords, comments, strings]

\newlength{\cslhangindent}
\newenvironment{cslreferences}%
  {}%
  {\par}
\usepackage{lmodern}
\usepackage{amssymb,amsmath}
\usepackage{ifxetex,ifluatex}
\usepackage{fixltx2e} 
\ifnum 0\ifxetex 1\fi\ifluatex 1\fi=0 
  \usepackage[T1]{fontenc}
  \usepackage[utf8]{inputenc}
\else 
  \ifxetex
    \usepackage{mathspec}
  \else
    \usepackage{fontspec}
  \fi
  \defaultfontfeatures{Ligatures=TeX,Scale=MatchLowercase}
\fi
\IfFileExists{upquote.sty}{\usepackage{upquote}}{}
\IfFileExists{microtype.sty}{%
\usepackage{microtype}
\UseMicrotypeSet[protrusion]{basicmath} 
}{}
\usepackage[unicode=true]{hyperref}
\hypersetup{
            pdftitle={The Interblockchain Communication Protocol: An Overview},
            pdfkeywords={ibc, interblockchain, dlt},
            pdfborder={0 0 0},
            breaklinks=true}
\ifnum 0\ifxetex 1\fi\ifluatex 1\fi=0 
  \usepackage[shorthands=off,main=english]{babel}
\else
  \usepackage{polyglossia}
  \setmainlanguage[]{english}
\fi
\usepackage{color}
\usepackage{fancyvrb}

\DefineVerbatimEnvironment{Highlighting}{Verbatim}{commandchars=\\\{\}}
\newenvironment{Shaded}{}{}

\newcommand{\AttributeTok}[1]{\textcolor[rgb]{0.49,0.56,0.16}{#1}}

\newcommand{\DataTypeTok}[1]{\textcolor[rgb]{0.56,0.13,0.00}{#1}}
\newcommand{\DecValTok}[1]{\textcolor[rgb]{0.25,0.63,0.44}{#1}}

\newcommand{\FunctionTok}[1]{\textcolor[rgb]{0.02,0.16,0.49}{#1}}

\newcommand{\KeywordTok}[1]{\textcolor[rgb]{0.00,0.44,0.13}{\textbf{#1}}}
\newcommand{\NormalTok}[1]{#1}
\newcommand{\OperatorTok}[1]{\textcolor[rgb]{0.40,0.40,0.40}{#1}}

\usepackage{longtable,booktabs}
\usepackage{supertabular}

\let\endhead\relax
\IfFileExists{footnote.sty}{\usepackage{footnote}\makesavenoteenv{long table}}{}
\IfFileExists{parskip.sty}{%
\usepackage{parskip}
}{
\setlength{\parindent}{0pt}
\setlength{\parskip}{6pt plus 2pt minus 1pt}
}
\providecommand{\tightlist}{%
  \setlength{\itemsep}{0pt}\setlength{\parskip}{0pt}}
\ifx\paragraph\undefined\else
\let\oldparagraph\paragraph
\renewcommand{\paragraph}[1]{\oldparagraph{#1}\mbox{}}
\fi
\ifx\subparagraph\undefined\else
\let\oldsubparagraph\subparagraph
\renewcommand{\subparagraph}[1]{\oldsubparagraph{#1}\mbox{}}
\fi

\makeatletter
\def\fps@figure{htbp}
\makeatother

\newcommand*{\TitleFont}{%
      \usefont{\encodingdefault}{\rmdefault}{b}{n}%
      \fontsize{18}{20}%
      \selectfont}

\title{\TitleFont The Interblockchain Communication Protocol: An
Overview}

\author{
            \IEEEauthorblockN{Christopher Goes}
        \IEEEauthorblockA{%
            Interchain GmbH \\
            Berlin, Germany \\
            cwgoes@interchain.berlin}
        }

\date{}

\begin{document}
\maketitle
\begin{abstract}
The interblockchain communication protocol (IBC) is an end-to-end,
connection-oriented, stateful protocol for reliable, ordered, and
authenticated communication between modules on separate distributed
ledgers. IBC is designed for interoperation between heterogenous ledgers
arranged in an unknown, dynamic topology, operating with varied
consensus algorithms and state machines. The protocol realises this by
specifying the sufficient set of data structures, abstractions, and
semantics of a communication protocol which once implemented by
participating ledgers will allow them to safely communicate. IBC is
payload-agnostic and provides a cross-ledger asynchronous communication
primitive which can be used as a constituent building block by a wide
variety of applications.
\end{abstract}


\hypertarget{introduction}{%
\section{Introduction}\label{introduction}}

By virtue of their nature as replicated state machines across which
deterministic execution and thus continued agreement on an exact
deterministic ruleset must be maintained, individual distributed ledgers
are limited in their throughput \& flexibility, must trade off
application-specific optimisations for general-purpose capabilities, and
can only offer a single security model to applications built on top of
them. In order to support the transaction throughput, application
diversity, cost efficiency, and fault tolerance required to facilitate
wide deployment of distributed ledger applications, execution and
storage must be split across many independent ledgers which can run
concurrently, upgrade independently, and each specialise in different
ways, in a manner such that the ability of different applications to
communicate with one another, essential for permissionless innovation
and complex multi-part contracts, is maintained.

One multi-ledger design direction is to shard a single logical ledger
across separate consensus instances, referred to as ``shards'', which
execute concurrently and store disjoint partitions of the state. In
order to reason globally about safety and liveness, and in order to
correctly route data and code between shards, these designs must take a
``top-down approach'' --- constructing a particular network topology,
usually a single root ledger and a star or tree of shards, and
engineering protocol rules and incentives to enforce that topology.
Message passing can then be implemented on top of such a sharded
topology by systems such as Polkadot's XCMP {[}1{]} and Ethereum 2.0's
cross-shard communication {[}2{]}. This approach possesses advantages in
simplicity and predictability, but faces hard technical problems in
assuring the validity of state transitions {[}3{]}, requires the
adherence of all shards to a single validator set (or randomly elected
subset thereof) and a single virtual machine, and faces challenges in
upgrading itself over time due to the necessity of reaching global
consensus on alterations to the network topology or ledger ruleset.
Additionally, such sharded systems are brittle: if the fault tolerance
threshold is exceeded, the system needs to coordinate a global halt \&
restart, and possibly initiate complex state transition rollback
procedures --- it is not possible to safely isolate Byzantine portions
of the network graph and continue operation.

The \emph{interblockchain communication protocol} (IBC) provides a
mechanism by which separate, sovereign replicated ledgers can safely,
voluntarily interact while sharing only a minimum requisite common
interface. The protocol design approaches a differently formulated
version of the scaling and interoperability problem: enabling safe,
reliable interoperation of a network of heterogeneous distributed
ledgers, arranged in an unknown topology, preserving data secrecy where
possible, where the ledgers can diversify, develop, and rearrange
independently of each other or of a particular imposed topology or
ledger design. In a wide, dynamic network of interoperating ledgers,
sporadic Byzantine faults are expected, so the protocol must also
detect, mitigate, and contain the potential damage of Byzantine faults
in accordance with the requirements of the applications and ledgers
involved without requiring the use of additional trusted parties or
global coordination.

To facilitate this heterogeneous interoperation, the interblockchain
communication protocol utilises a bottom-up approach, specifying the set
of requirements, functions, and properties necessary to implement
interoperation between two ledgers, and then specifying different ways
in which multiple interoperating ledgers might be composed which
preserve the requirements of higher-level protocols. IBC thus presumes
nothing about and requires nothing of the overall network topology, and
of the implementing ledgers requires only that a known, minimal set of
functions with specified properties are available. Ledgers within IBC
are defined as their light client consensus validation functions, thus
expanding the range of what a ``ledger'' can be to include single
machines and complex consensus algorithms alike. IBC implementations are
expected to be co-resident with higher-level modules and protocols on
the host ledger. Ledgers hosting IBC must provide a certain set of
functions for consensus transcript verification and cryptographic
commitment proof generation, and IBC packet relayers (off-ledger
processes) are expected to have access to network protocols and physical
data-links as required to read the state of one ledger and submit data
to another.

The data payloads in IBC packets are opaque to the protocol itself ---
modules on each ledger determine the semantics of the packets which are
sent between them. For cross-ledger token transfer, packets could
contain fungible token information, where assets are locked on one
ledger to mint corresponding vouchers on another. For cross-ledger
governance, packets could contain vote information, where accounts on
one ledger could vote in the governance system of another. For
cross-ledger account delegation, packets could contain transaction
authorisation information, allowing an account on one ledger to be
controlled by an account on another. For a cross-ledger decentralised
exchange, packets could contain order intent information or trade
settlement information, such that assets on different ledgers could be
exchanged without leaving their host ledgers by transitory escrow and a
sequence of packets.

This bottom-up approach is quite similar to, and directly inspired by,
the TCP/IP specification {[}4{]} for interoperability between hosts in
packet-switched computer networks. Just as TCP/IP defines the protocol
by which two hosts communicate, and higher-level protocols knit many
bidirectional host-to-host links into complex topologies, IBC defines
the protocol by which two ledgers communicate, and higher-level
protocols knit many bidirectional ledger-to-ledger links into gestalt
multi-ledger applications. Just as TCP/IP packets contain opaque payload
data with semantics interpreted by the processes on each host, IBC
packets contain opaque payload data with semantics interpreted by the
modules on each ledger. Just as TCP/IP provides reliable, ordered data
transmission between processes, allowing a process on one host to reason
about the state of a process on another, IBC provides reliable, ordered
data transmission between modules, allowing a module on one ledger to
reason about the state of a module on another.

This paper is intended as an overview of the abstractions defined by the
IBC protocol and the mechanisms by which they are composed. We first
outline the structure of the protocol, including scope, interfaces, and
operational requirements. Subsequently, we detail the abstractions
defined by the protocol, including modules, ports, clients, connections,
channels, packets, and relayers, and describe the subprotocols for
opening and closing handshakes, packet relay, edge-case handling, and
relayer operations. After explaining the internal structure of the
protocol, we define the interface by which applications can utilise IBC,
and sketch an example application-level protocol for fungible token
transfer. Finally, we recount testing and deployment efforts of the
protocol thus far. Appendices include pseudocode for the connection
handshake, channel handshake, and packet relay algorithms.

\hypertarget{protocol-scope-properties}{%
\section{Protocol scope \& properties}\label{protocol-scope-properties}}

\hypertarget{scope}{%
\subsection{Scope}\label{scope}}

IBC handles authentication, transport, and ordering of opaque data
packets relayed between modules on separate ledgers --- ledgers can be
run on solo machines, replicated by many nodes running a consensus
algorithm, or constructed by any process whose state can be verified.
The protocol is defined between modules on two ledgers, but designed for
safe simultaneous use between any number of modules on any number of
ledgers connected in arbitrary topologies.

\hypertarget{interfaces}{%
\subsection{Interfaces}\label{interfaces}}

IBC sits between modules --- smart contracts, other ledger components,
or otherwise independently executed pieces of application logic on
ledgers --- on one side, and underlying consensus protocols,
blockchains, and network infrastructure (e.g.~TCP/IP), on the other
side.

IBC provides to modules a set of functions much like the functions which
might be provided to a module for interacting with another module on the
same ledger: sending data packets and receiving data packets on an
established connection and channel, in addition to calls to manage the
protocol state: opening and closing connections and channels, choosing
connection, channel, and packet delivery options, and inspecting
connection and channel status.

IBC requires certain functionalities and properties of the underlying
ledgers, primarily finality (or thresholding finality gadgets),
cheaply-verifiable consensus transcripts (such that a light client
algorithm can verify the results of the consensus process with much less
computation \& storage than a full node), and simple key/value store
functionality. On the network side, IBC requires only eventual data
delivery --- no authentication, synchrony, or ordering properties are
assumed.

\hypertarget{operation}{%
\subsection{Operation}\label{operation}}

The primary purpose of IBC is to provide reliable, authenticated,
ordered communication between modules running on independent host
ledgers. This requires protocol logic in the areas of data relay, data
confidentiality and legibility, reliability, flow control,
authentication, statefulness, and multiplexing.

\vspace{3mm}

\hypertarget{data-relay}{%
\subsubsection{Data relay}\label{data-relay}}

\vspace{3mm}

In the IBC architecture, modules are not directly sending messages to
each other over networking infrastructure, but rather are creating
messages to be sent which are then physically relayed from one ledger to
another by monitoring ``relayer processes''. IBC assumes the existence
of a set of relayer processes with access to an underlying network
protocol stack (likely TCP/IP, UDP/IP, or QUIC/IP) and physical
interconnect infrastructure. These relayer processes monitor a set of
ledgers implementing the IBC protocol, continuously scanning the state
of each ledger and requesting transaction execution on another ledger
when outgoing packets have been committed. For correct operation and
progress in a connection between two ledgers, IBC requires only that at
least one correct and live relayer process exists which can relay
between the ledgers.

\vspace{3mm}

\hypertarget{data-confidentiality-and-legibility}{%
\subsubsection{Data confidentiality and
legibility}\label{data-confidentiality-and-legibility}}

\vspace{3mm}

The IBC protocol requires only that the minimum data necessary for
correct operation of the IBC protocol be made available and legible
(serialised in a standardised format) to relayer processes, and the
ledger may elect to make that data available only to specific relayers.
This data consists of consensus state, client, connection, channel, and
packet information, and any auxiliary state structure necessary to
construct proofs of inclusion or exclusion of particular key/value pairs
in state. All data which must be proved to another ledger must also be
legible; i.e., it must be serialised in a standardised format agreed
upon by the two ledgers.

\vspace{3mm}

\hypertarget{reliability}{%
\subsubsection{Reliability}\label{reliability}}

\vspace{3mm}

The network layer and relayer processes may behave in arbitrary ways,
dropping, reordering, or duplicating packets, purposely attempting to
send invalid transactions, or otherwise acting in a Byzantine fashion,
without compromising the safety or liveness of IBC. This is achieved by
assigning a sequence number to each packet sent over an IBC channel,
which is checked by the IBC handler (the part of the ledger implementing
the IBC protocol) on the receiving ledger, and providing a method for
the sending ledger to check that the receiving ledger has in fact
received and handled a packet before sending more packets or taking
further action. Cryptographic commitments are used to prevent datagram
forgery: the sending ledger commits to outgoing packets, and the
receiving ledger checks these commitments, so datagrams altered in
transit by a relayer will be rejected. IBC also supports unordered
channels, which do not enforce ordering of packet receives relative to
sends but still enforce exactly-once delivery.

\vspace{3mm}

\hypertarget{flow-control}{%
\subsubsection{Flow control}\label{flow-control}}

\vspace{3mm}

IBC does not provide specific protocol-level provisions for
compute-level or economic-level flow control. The underlying ledgers are
expected to have compute throughput limiting devices and flow control
mechanisms of their own such as gas markets. Application-level economic
flow control --- limiting the rate of particular packets according to
their content --- may be useful to ensure security properties and
contain damage from Byzantine faults. For example, an application
transferring value over an IBC channel might want to limit the rate of
value transfer per block to limit damage from potential Byzantine
behaviour. IBC provides facilities for modules to reject packets and
leaves particulars up to the higher-level application protocols.

\vspace{3mm}

\hypertarget{authentication}{%
\subsubsection{Authentication}\label{authentication}}

\vspace{3mm}

All data sent over IBC are authenticated: a block finalised by the
consensus algorithm of the sending ledger must commit to the outgoing
packet via a cryptographic commitment, and the receiving ledger's IBC
handler must verify both the consensus transcript and the cryptographic
commitment proof that the datagram was sent before acting upon it.

\vspace{3mm}

\hypertarget{statefulness}{%
\subsubsection{Statefulness}\label{statefulness}}

\vspace{3mm}

Reliability, flow control, and authentication as described above require
that IBC initialises and maintains certain status information for each
datastream. This information is split between three abstractions:
clients, connections, and channels. Each client object contains
information about the consensus state of the counterparty ledger. Each
connection object contains a specific pair of named identifiers agreed
to by both ledgers in a handshake protocol, which uniquely identifies a
connection between the two ledgers. Each channel, specific to a pair of
modules, contains information concerning negotiated encoding and
multiplexing options and state and sequence numbers. When two modules
wish to communicate, they must locate an existing connection and channel
between their two ledgers, or initialise a new connection and channel(s)
if none yet exist. Initialising connections and channels requires a
multi-step handshake which, once complete, ensures that only the two
intended ledgers are connected, in the case of connections, and ensures
that two modules are connected and that future datagrams relayed will be
authenticated, encoded, and sequenced as desired, in the case of
channels.

\vspace{3mm}

\hypertarget{multiplexing}{%
\subsubsection{Multiplexing}\label{multiplexing}}

\vspace{3mm}

To allow for many modules within a single host ledger to use an IBC
connection simultaneously, IBC allows any number of channels to be
associated with a single connection. Each channel uniquely identifies a
datastream over which packets can be sent in order (in the case of an
ordered channel), and always exactly once, to a destination module on
the receiving ledger. Channels are usually expected to be associated
with a single module on each ledger, but one-to-many and many-to-one
channels are also possible. The number of channels per connection is
unbounded, facilitating concurrent throughput limited only by the
throughput of the underlying ledgers with only a single connection and
pair of clients necessary to track consensus information (and consensus
transcript verification cost thus amortised across all channels using
the connection).

\hypertarget{host-ledger-requirements}{%
\section{Host ledger requirements}\label{host-ledger-requirements}}

\hypertarget{module-system}{%
\subsubsection{Module system}\label{module-system}}

\vspace{3mm}

The host ledger must support a module system, whereby self-contained,
potentially mutually distrusted packages of code can safely execute on
the same ledger, control how and when they allow other modules to
communicate with them, and be identified and manipulated by a controller
module or execution environment.

\vspace{3mm}

\hypertarget{keyvalue-store}{%
\subsubsection{Key/value Store}\label{keyvalue-store}}

\vspace{3mm}

The host ledger must provide a key/value store interface allowing values
to be read, written, and deleted.

These functions must be permissioned to the IBC handler module so that
only the IBC handler module can write or delete a certain subset of
paths. This will likely be implemented as a sub-store (prefixed
key-space) of a larger key/value store used by the entire ledger.

Host ledgers must provide an instance of this interface which is
provable, such that the light client algorithm for the host ledger can
verify presence or absence of particular key-value pairs which have been
written to it.

This interface does not necessitate any particular storage backend or
backend data layout. ledgers may elect to use a storage backend
configured in accordance with their needs, as long as the store on top
fulfils the specified interface and provides commitment proofs.

\vspace{3mm}

\hypertarget{consensus-state-introspection}{%
\subsubsection{Consensus state
introspection}\label{consensus-state-introspection}}

\vspace{3mm}

Host ledgers must provide the ability to introspect their current
height, current consensus state (as utilised by the host ledger's light
client algorithm), and a bounded number of recent consensus states
(e.g.~past headers). These are used to prevent man-in-the-middle attacks
during handshakes to set up connections with other ledgers --- each
ledger checks that the other ledger is in fact authenticating data using
its consensus state.

\vspace{3mm}

\hypertarget{timestamp-access}{%
\subsubsection{Timestamp access}\label{timestamp-access}}

\vspace{3mm}

In order to support timestamp-based timeouts, host ledgers must provide
a current Unix-style timestamp. Timeouts in subsequent headers must be
non-decreasing.

\vspace{3mm}

\hypertarget{port-system}{%
\subsubsection{Port system}\label{port-system}}

\vspace{3mm}

Host ledgers must implement a port system, where the IBC handler can
allow different modules in the host ledger to bind to uniquely named
ports. Ports are identified by an identifier, and must be permissioned
so that:

\begin{itemize}
\tightlist
\item
  Once a module has bound to a port, no other modules can use that port
  until the module releases it
\item
  A single module can bind to multiple ports
\item
  Ports are allocated first-come first-serve
\item
  ``Reserved'' ports for known modules can be bound when the ledger is
  first started
\end{itemize}

This permissioning can be implemented with unique references (object
capabilities {[}5{]}) for each port, with source-based authentication(a
la \texttt{msg.sender} in Ethereum contracts), or with some other method
of access control, in any case enforced by the host ledger.

Ports are not generally intended to be human-readable identifiers ---
just as DNS name resolution and standardised port numbers for particular
applications exist to abstract away the details of IP addresses and
ports from TCP/IP users, ledger name resolution and standardised ports
for particular applications may be created in order to abstract away the
details of ledger identification and port selection. Such an addressing
system could easily be built on top of IBC itself, such that an initial
connection to the addressing system over IBC would then enable name
resolution for subsequent connections to other ledgers and applications.

\vspace{3mm}

\hypertarget{exceptionrollback-system}{%
\subsubsection{Exception/rollback
system}\label{exceptionrollback-system}}

\vspace{3mm}

Host ledgers must support an exception or rollback system, whereby a
transaction can abort execution and revert any previously made state
changes (including state changes in other modules happening within the
same transaction), excluding gas consumed and fee payments as
appropriate.

\vspace{3mm}

\hypertarget{data-availability}{%
\subsubsection{Data availability}\label{data-availability}}

\vspace{3mm}

For deliver-or-timeout safety, host ledgers must have eventual data
availability, such that any key/value pairs in state can be eventually
retrieved by relayers. For exactly-once safety, data availability is not
required.

For liveness of packet relay, host ledgers must have bounded
transactional liveness, such that incoming transactions are confirmed
within a block height or timestamp bound (in particular, less than the
timeouts assigned to the packets).

IBC packet data, and other data which is not directly stored in the
Merklized state but is relied upon by relayers, must be available to and
efficiently computable by relayer processes.

\hypertarget{protocol-structure}{%
\section{Protocol structure}\label{protocol-structure}}

\hypertarget{clients}{%
\subsection{Clients}\label{clients}}

The \emph{client} abstraction encapsulates the properties that consensus
algorithms of ledgers implementing the interblockchain communication
protocol are required to satisfy. These properties are necessary for
efficient and safe state verification in the higher-level protocol
abstractions. The algorithm utilised in IBC to verify the consensus
transcript and state sub-components of another ledger is referred to as
a ``validity predicate'', and pairing it with a state that the verifier
assumes to be correct forms a ``light client'' (colloquially shortened
to ``client'').

\vspace{3mm}

\hypertarget{motivation}{%
\subsubsection{Motivation}\label{motivation}}

\vspace{3mm}

In the IBC protocol, an actor, which may be an end user, an off-ledger
process, or ledger, needs to be able to verify updates to the state of
another ledger which the other ledger's consensus algorithm has agreed
upon, and reject any possible updates which the other ledger's consensus
algorithm has not agreed upon. A light client is the algorithm with
which an actor can do so. The client abstraction formalises this model's
interface and requirements, so that the IBC protocol can easily
integrate with new ledgers which are running new consensus algorithms as
long as associated light client algorithms fulfilling the listed
requirements are provided.

Beyond the properties described in this specification, IBC does not
impose any requirements on the internal operation of ledgers and their
consensus algorithms. A ledger may consist of a single process signing
operations with a private key, a quorum of processes signing in unison,
many processes operating a Byzantine fault-tolerant consensus algorithm
(a replicated, or distributed, ledger), or other configurations yet to
be invented --- from the perspective of IBC, a ledger is defined
entirely by its light client validation and equivocation detection
logic. Clients will generally not include validation of the state
transition logic in general (as that would be equivalent to simply
executing the other state machine), but may elect to validate parts of
state transitions in particular cases, and can validate the entire state
transition if doing so is asymptotically efficient, perhaps through
compression using a SNARK {[}6{]}.

Externally, however, the light client verification functions used by IBC
clients must have \emph{finality}, such that verified blocks (subject to
the usual consensus safety assumptions), once verified, cannot be
reverted. The safety of higher abstraction layers of the IBC protocol
and guarantees provided to the applications using the protocol depend on
this property of finality.

In order to graft finality onto Nakamoto consensus algorithms, such as
used in Bitcoin {[}7{]}, clients can act as thresholding views of
internal, non-finalising clients. In the case where modules utilising
the IBC protocol to interact with probabilistic-finality consensus
algorithms which might require different finality thresholds for
different applications, one write-only client could be created to track
headers and many read-only clients with different finality thresholds
(confirmation depths after which state roots are considered final) could
use that same state. Of course, this will introduce different security
assumptions than those required of full nodes running the consensus
algorithm, and trade-offs which must be balanced by the user on the
basis of their application-specific security needs.

The client protocol is designed to support third-party introduction.
Consider the general example: Alice, a module on a ledger, wants to
introduce Bob, a second module on a second ledger who Alice knows (and
who knows Alice), to Carol, a third module on a third ledger, who Alice
knows but Bob does not. Alice must utilise an existing channel to Bob to
communicate the canonically-serialisable validity predicate for Carol,
with which Bob can then open a connection and channel so that Bob and
Carol can talk directly. If necessary, Alice may also communicate to
Carol the validity predicate for Bob, prior to Bob's connection attempt,
so that Carol knows to accept the incoming request.

Client interfaces are constructed so that custom validation logic can be
provided safely to define a custom client at runtime, as long as the
underlying ledger can provide an appropriate gas metering mechanism to
charge for compute and storage. On a host ledger which supports WASM
execution, for example, the validity predicate and equivocation
predicate could be provided as executable WASM functions when the client
instance is created.

\vspace{3mm}

\hypertarget{definitions}{%
\subsubsection{Definitions}\label{definitions}}

\vspace{3mm}

A \emph{validity predicate} is an opaque function defined by a client
type to verify headers depending on the current consensus state. Using
the validity predicate should be far more computationally efficient than
replaying the full consensus algorithm and state machine for the given
parent header and the list of network messages.

A \emph{consensus state} is an opaque type representing the state of a
validity predicate. The light client validity predicate algorithm in
combination with a particular consensus state must be able to verify
state updates agreed upon by the associated consensus algorithm. The
consensus state must also be serialisable in a canonical fashion so that
third parties, such as counterparty ledgers, can check that a particular
ledger has stored a particular state. It must also be introspectable by
the ledger which it is for, such that the ledger can look up its own
consensus state at a past height and compare it to a stored consensus
state in another ledger's client.

A \emph{commitment root} is an inexpensive way for downstream logic to
verify whether key/value pairs are present or absent in a state at a
particular height. Often this will be instantiated as the root of a
Merkle tree.

A \emph{header} is an opaque data structure defined by a client type
which provides information to update a consensus state. Headers can be
submitted to an associated client to update the stored consensus state.
They likely contain a height, a proof, a new commitment root, and
possibly updates to the validity predicate.

A \emph{misbehaviour predicate} is an opaque function defined by a
client type, used to check if data constitutes a violation of the
consensus protocol. This might be two signed headers with different
state roots but the same height, a signed header containing invalid
state transitions, or other evidence of malfeasance as defined by the
consensus algorithm.

\vspace{3mm}

\hypertarget{desired-properties}{%
\subsubsection{Desired properties}\label{desired-properties}}

\vspace{3mm}

Light clients must provide a secure algorithm to verify other ledgers'
canonical headers, using the existing consensus state. The higher level
abstractions will then be able to verify sub-components of the state
with the commitment roots stored in the consensus state, which are
guaranteed to have been committed by the other ledger's consensus
algorithm.

Validity predicates are expected to reflect the behaviour of the full
nodes which are running the corresponding consensus algorithm. Given a
consensus state and a list of messages, if a full node accepts a new
header, then the light client must also accept it, and if a full node
rejects it, then the light client must also reject it.

Light clients are not replaying the whole message transcript, so it is
possible under cases of consensus misbehaviour that the light clients'
behaviour differs from the full nodes'. In this case, a misbehaviour
proof which proves the divergence between the validity predicate and the
full node can be generated and submitted to the ledger so that the
ledger can safely deactivate the light client, invalidate past state
roots, and await higher-level intervention.

The validity of the validity predicate is dependent on the security
model of the consensus algorithm. For example, the consensus algorithm
could be BFT proof-of-authority with a trusted operator set, or BFT
proof-of-stake with a tokenholder set, each of which have a defined
threshold above which Byzantine behaviour may result in divergence.

Clients may have time-sensitive validity predicates, such that if no
header is provided for a period of time (e.g.~an unbonding period of
three weeks in a proof-of-stake system) it will no longer be possible to
update the client.

\vspace{3mm}

\hypertarget{state-verification}{%
\subsubsection{State verification}\label{state-verification}}

\vspace{3mm}

Client types must define functions to authenticate internal state of the
ledger which the client tracks. Internal implementation details may
differ (for example, a loopback client could simply read directly from
the state and require no proofs). Externally-facing clients will likely
verify signature or vector commitment proofs.

\vspace{3mm}

\hypertarget{example-client-instantiations}{%
\subsubsection{Example client
instantiations}\label{example-client-instantiations}}

\vspace{3mm}

\hypertarget{loopback}{%
\paragraph{Loopback}\label{loopback}}

A loopback client of a local ledger merely reads from the local state,
to which it must have access. This is analogous to \texttt{localhost} or
\texttt{127.0.0.1} in TCP/IP.

\vspace{3mm}

\hypertarget{simple-signatures}{%
\paragraph{Simple signatures}\label{simple-signatures}}

A client of a solo machine running a non-replicated ledger with a known
public key checks signatures on messages sent by that local machine.
Multi-signature or threshold signature schemes can also be used in such
a fashion.

\vspace{3mm}

\hypertarget{proxy-clients}{%
\paragraph{Proxy clients}\label{proxy-clients}}

Proxy clients verify another (proxy) ledger's verification of the target
ledger, by including in the proof first a proof of the client state on
the proxy ledger, and then a secondary proof of the sub-state of the
target ledger with respect to the client state on the proxy ledger. This
allows the proxy client to avoid storing and tracking the consensus
state of the target ledger itself, at the cost of adding security
assumptions of proxy ledger correctness.

\vspace{3mm}

\hypertarget{bft-consensus-and-verifiable-state}{%
\paragraph{BFT consensus and verifiable
state}\label{bft-consensus-and-verifiable-state}}

For the immediate application of interoperability between sovereign,
fault-tolerant distributed ledgers, the most common and most useful
client type will be light clients for instances of BFT consensus
algorithms such as Tendermint {[}8{]}, GRANDPA {[}9{]}, or HotStuff
{[}10{]}, with ledgers utilising Merklized state trees such as an IAVL+
tree {[}11{]} or a Merkle Patricia tree {[}12{]}. The client algorithm
for such instances will utilise the BFT consensus algorithm's light
client validity predicate and treat at minimum consensus equivocation
(double-signing) as misbehaviour, along with other possible misbehaviour
types specific to the proof-of-authority or proof-of-stake system
involved.

\vspace{3mm}

\hypertarget{client-lifecycle}{%
\subsubsection{Client lifecycle}\label{client-lifecycle}}

\vspace{3mm}

\hypertarget{creation}{%
\paragraph{Creation}\label{creation}}

Clients can be created permissionlessly by anyone at any time by
specifying an identifier, client type, and initial consensus state.

\vspace{3mm}

\hypertarget{update}{%
\paragraph{Update}\label{update}}

Updating a client is done by submitting a new header. When a new header
is verified with the stored client state's validity predicate and
consensus state, the client will update its internal state accordingly,
possibly finalising commitment roots and updating the signature
authority logic in the stored consensus state.

If a client can no longer be updated (if, for example, the unbonding
period has passed), it will no longer be possible to send any packets
over connections and channels associated with that client, or timeout
any packets in-flight (since the height and timestamp on the destination
ledger can no longer be verified). Manual intervention must take place
to reset the client state or migrate the connections and channels to
another client. This cannot safely be done automatically, but ledgers
implementing IBC could elect to allow governance mechanisms to perform
these actions (perhaps even per-client/connection/channel with a
controlling multi-signature or contract).

\vspace{3mm}

\hypertarget{misbehaviour}{%
\paragraph{Misbehaviour}\label{misbehaviour}}

If the client detects evidence of misbehaviour, the client can be take
appropriate action, possibly invalidating previously valid commitment
roots and preventing future updates. What precisely constitutes
misbehaviour will depend on the consensus algorithm which the validity
predicate is validating the output of.

\hypertarget{connections}{%
\subsection{Connections}\label{connections}}

The \emph{connection} abstraction encapsulates two stateful objects
(\emph{connection ends}) on two separate ledgers, each associated with a
light client of the other ledger, which together facilitate cross-ledger
sub-state verification and packet relay (through channels). Connections
are safely established in an unknown, dynamic topology using a handshake
subprotocol.

\vspace{3mm}

\hypertarget{motivation}{%
\subsubsection{Motivation}\label{motivation}}

\vspace{3mm}

The IBC protocol provides \emph{authorisation} and \emph{ordering}
semantics for packets: guarantees, respectively, that packets have been
committed on the sending ledger (and according state transitions
executed, such as escrowing tokens), and that they have been committed
exactly once in a particular order and can be delivered exactly once in
that same order. The \emph{connection} abstraction in conjunction with
the \emph{client} abstraction defines the \emph{authorisation} semantics
of IBC. Ordering semantics are provided by channels.

\vspace{3mm}

\hypertarget{definitions}{%
\subsubsection{Definitions}\label{definitions}}

\vspace{3mm}

A \emph{connection end} is state tracked for an end of a connection on
one ledger, defined as follows:

\begin{Shaded}
\begin{Highlighting}[]
\NormalTok{enum ConnectionState \{}
\NormalTok{  INIT}\OperatorTok{,}
\NormalTok{  TRYOPEN}\OperatorTok{,}
\NormalTok{  OPEN}\OperatorTok{,}
\NormalTok{\}}
\end{Highlighting}
\end{Shaded}

\begin{Shaded}
\begin{Highlighting}[]
\NormalTok{interface ConnectionEnd \{}
\NormalTok{  state}\OperatorTok{:}\NormalTok{ ConnectionState}
\NormalTok{  counterpartyConnectionIdentifier}\OperatorTok{:}\NormalTok{ Identifier}
\NormalTok{  counterpartyPrefix}\OperatorTok{:}\NormalTok{ CommitmentPrefix}
\NormalTok{  clientIdentifier}\OperatorTok{:}\NormalTok{ Identifier}
\NormalTok{  counterpartyClientIdentifier}\OperatorTok{:}\NormalTok{ Identifier}
\NormalTok{  version}\OperatorTok{:} \DataTypeTok{string}
\NormalTok{\}}
\end{Highlighting}
\end{Shaded}

\begin{itemize}
\tightlist
\item
  The \texttt{state} field describes the current state of the connection
  end.
\item
  The \texttt{counterpartyConnectionIdentifier} field identifies the
  connection end on the counterparty ledger associated with this
  connection.
\item
  The \texttt{counterpartyPrefix} field contains the prefix used for
  state verification on the counterparty ledger associated with this
  connection.
\item
  The \texttt{clientIdentifier} field identifies the client associated
  with this connection.
\item
  The \texttt{counterpartyClientIdentifier} field identifies the client
  on the counterparty ledger associated with this connection.
\item
  The \texttt{version} field is an opaque string which can be utilised
  to determine encodings or protocols for channels or packets utilising
  this connection.
\end{itemize}

\vspace{3mm}

\hypertarget{opening-handshake}{%
\subsubsection{Opening handshake}\label{opening-handshake}}

\vspace{3mm}

The opening handshake subprotocol allows each ledger to verify the
identifier used to reference the connection on the other ledger,
enabling modules on each ledger to reason about the reference on the
other ledger.

The opening handshake consists of four datagrams: \texttt{ConnOpenInit},
\texttt{ConnOpenTry}, \texttt{ConnOpenAck}, and
\texttt{ConnOpenConfirm}.

A correct protocol execution, between two ledgers \texttt{A} and
\texttt{B}, with connection states formatted as \texttt{(A,\ B)}, flows
as follows:

\begin{longtable}[]{@{}lll@{}}
\toprule
Datagram & Prior state & Posterior state\tabularnewline
\midrule
\endhead
\texttt{ConnOpenInit} & \texttt{(-,\ -)} &
\texttt{(INIT,\ -)}\tabularnewline
\texttt{ConnOpenTry} & \texttt{(INIT,\ none)} &
\texttt{(INIT,\ TRYOPEN)}\tabularnewline
\texttt{ConnOpenAck} & \texttt{(INIT,\ TRYOPEN)} &
\texttt{(OPEN,\ TRYOPEN)}\tabularnewline
\texttt{ConnOpenConfirm} & \texttt{(OPEN,\ TRYOPEN)} &
\texttt{(OPEN,\ OPEN)}\tabularnewline
\bottomrule
\end{longtable}

At the end of an opening handshake between two ledgers implementing the
subprotocol, the following properties hold:

\begin{itemize}
\tightlist
\item
  Each ledger has each other's correct consensus state as originally
  specified by the initiating actor.
\item
  Each ledger has knowledge of and has agreed to its identifier on the
  other ledger.
\item
  Each ledger knows that the other ledger has agreed to the same data.
\end{itemize}

Connection handshakes can safely be performed permissionlessly, modulo
anti-spam measures (paying gas).

\texttt{ConnOpenInit}, executed on ledger A, initialises a connection
attempt on ledger A, specifying a pair of identifiers for the connection
on both ledgers and a pair of identifiers for existing light clients
(one for each ledger). ledger A stores a connection end object in its
state.

\texttt{ConnOpenTry}, executed on ledger B, relays notice of a
connection attempt on ledger A to ledger B, providing the pair of
connection identifiers, the pair of client identifiers, and a desired
version. Ledger B verifies that these identifiers are valid, checks that
the version is compatible, verifies a proof that ledger A has stored
these identifiers, and verifies a proof that the light client ledger A
is using to validate ledger B has the correct consensus state for ledger
B. ledger B stores a connection end object in its state.

\texttt{ConnOpenAck}, executed on ledger A, relays acceptance of a
connection open attempt from ledger B back to ledger A, providing the
identifier which can now be used to look up the connection end object.
ledger A verifies that the version requested is compatible, verifies a
proof that ledger B has stored the same identifiers ledger A has stored,
and verifies a proof that the light client ledger B is using to validate
ledger A has the correct consensus state for ledger A.

\texttt{ConnOpenConfirm}, executed on ledger B, confirms opening of a
connection on ledger A to ledger B. Ledger B simply checks that ledger A
has executed \texttt{ConnOpenAck} and marked the connection as
\texttt{OPEN}. Ledger B subsequently marks its end of the connection as
\texttt{OPEN}. After execution of \texttt{ConnOpenConfirm} the
connection is open on both ends and can be used immediately.

\vspace{3mm}

\hypertarget{versioning}{%
\subsubsection{Versioning}\label{versioning}}

\vspace{3mm}

During the handshake process, two ends of a connection come to agreement
on a version bytestring associated with that connection. At the moment,
the contents of this version bytestring are opaque to the IBC core
protocol. In the future, it might be used to indicate what kinds of
channels can utilise the connection in question, or what encoding
formats channel-related datagrams will use. Host ledgers may utilise the
version data to negotiate encodings, priorities, or connection-specific
metadata related to custom logic on top of IBC. Host ledgers may also
safely ignore the version data or specify an empty string.

\hypertarget{channels}{%
\subsection{Channels}\label{channels}}

The \emph{channel} abstraction provides message delivery semantics to
the interblockchain communication protocol in three categories:
ordering, exactly-once delivery, and module permissioning. A channel
serves as a conduit for packets passing between a module on one ledger
and a module on another, ensuring that packets are executed only once,
delivered in the order in which they were sent (if necessary), and
delivered only to the corresponding module owning the other end of the
channel on the destination ledger. Each channel is associated with a
particular connection, and a connection may have any number of
associated channels, allowing the use of common identifiers and
amortising the cost of header verification across all the channels
utilising a connection and light client.

Channels are payload-agnostic. The modules which send and receive IBC
packets decide how to construct packet data and how to act upon the
incoming packet data, and must utilise their own application logic to
determine which state transactions to apply according to what data the
packet contains.

\vspace{3mm}

\hypertarget{motivation}{%
\subsubsection{Motivation}\label{motivation}}

\vspace{3mm}

The interblockchain communication protocol uses a cross-ledger message
passing model. IBC \emph{packets} are relayed from one ledger to the
other by external relayer processes. Two ledgers, A and B, confirm new
blocks independently, and packets from one ledger to the other may be
delayed, censored, or re-ordered arbitrarily. Packets are visible to
relayers and can be read from a ledger by any relayer process and
submitted to any other ledger.

The IBC protocol must provide ordering (for ordered channels) and
exactly-once delivery guarantees to allow applications to reason about
the combined state of connected modules on two ledgers. For example, an
application may wish to allow a single tokenised asset to be transferred
between and held on multiple ledgers while preserving fungibility and
conservation of supply. The application can mint asset vouchers on
ledger B when a particular IBC packet is committed to ledger B, and
require outgoing sends of that packet on ledger A to escrow an equal
amount of the asset on ledger A until the vouchers are later redeemed
back to ledger A with an IBC packet in the reverse direction. This
ordering guarantee along with correct application logic can ensure that
total supply is preserved across both ledgers and that any vouchers
minted on ledger B can later be redeemed back to ledger A. A more
detailed explanation of this example is provided later on.

\vspace{3mm}

\hypertarget{definitions}{%
\subsubsection{Definitions}\label{definitions}}

\vspace{3mm}

A \emph{channel} is a pipeline for exactly-once packet delivery between
specific modules on separate ledgers, which has at least one end capable
of sending packets and one end capable of receiving packets.

An \emph{ordered} channel is a channel where packets are delivered
exactly in the order which they were sent.

An \emph{unordered} channel is a channel where packets can be delivered
in any order, which may differ from the order in which they were sent.

All channels provide exactly-once packet delivery, meaning that a packet
sent on one end of a channel is delivered no more and no less than once,
eventually, to the other end.

A \emph{channel end} is a data structure storing metadata associated
with one end of a channel on one of the participating ledgers, defined
as follows:

\begin{Shaded}
\begin{Highlighting}[]
\NormalTok{interface ChannelEnd \{}
\NormalTok{  state}\OperatorTok{:}\NormalTok{ ChannelState}
\NormalTok{  ordering}\OperatorTok{:}\NormalTok{ ChannelOrder}
\NormalTok{  counterpartyPortIdentifier}\OperatorTok{:}\NormalTok{ Identifier}
\NormalTok{  counterpartyChannelIdentifier}\OperatorTok{:}\NormalTok{ Identifier}
\NormalTok{  nextSequenceSend}\OperatorTok{:}\NormalTok{ uint64}
\NormalTok{  nextSequenceRecv}\OperatorTok{:}\NormalTok{ uint64}
\NormalTok{  nextSequenceAck}\OperatorTok{:}\NormalTok{ uint64}
\NormalTok{  connectionHops}\OperatorTok{:}\NormalTok{ [Identifier]}
\NormalTok{  version}\OperatorTok{:} \DataTypeTok{string}
\NormalTok{\}}
\end{Highlighting}
\end{Shaded}

\begin{itemize}
\tightlist
\item
  The \texttt{state} is the current state of the channel end.
\item
  The \texttt{ordering} field indicates whether the channel is ordered
  or unordered. This is an enumeration instead of a boolean in order to
  allow additional kinds of ordering to be easily supported in the
  future.
\item
  The \texttt{counterpartyPortIdentifier} identifies the port on the
  counterparty ledger which owns the other end of the channel.
\item
  The \texttt{counterpartyChannelIdentifier} identifies the channel end
  on the counterparty ledger.
\item
  The \texttt{nextSequenceSend}, stored separately, tracks the sequence
  number for the next packet to be sent.
\item
  The \texttt{nextSequenceRecv}, stored separately, tracks the sequence
  number for the next packet to be received.
\item
  The \texttt{nextSequenceAck}, stored separately, tracks the sequence
  number for the next packet to be acknowledged.
\item
  The \texttt{connectionHops} stores the list of connection identifiers,
  in order, along which packets sent on this channel will travel. At the
  moment this list must be of length 1. In the future multi-hop channels
  may be supported.
\item
  The \texttt{version} string stores an opaque channel version, which is
  agreed upon during the handshake. This can determine module-level
  configuration such as which packet encoding is used for the channel.
  This version is not used by the core IBC protocol.
\end{itemize}

Channel ends have a \emph{state}:

\begin{Shaded}
\begin{Highlighting}[]
\NormalTok{enum ChannelState \{}
\NormalTok{  INIT}\OperatorTok{,}
\NormalTok{  TRYOPEN}\OperatorTok{,}
\NormalTok{  OPEN}\OperatorTok{,}
\NormalTok{  CLOSED}\OperatorTok{,}
\NormalTok{\}}
\end{Highlighting}
\end{Shaded}

\begin{itemize}
\tightlist
\item
  A channel end in \texttt{INIT} state has just started the opening
  handshake.
\item
  A channel end in \texttt{TRYOPEN} state has acknowledged the handshake
  step on the counterparty ledger.
\item
  A channel end in \texttt{OPEN} state has completed the handshake and
  is ready to send and receive packets.
\item
  A channel end in \texttt{CLOSED} state has been closed and can no
  longer be used to send or receive packets.
\end{itemize}

A \texttt{Packet}, encapsulating opaque data to be transferred from one
module to another over a channel, is a particular interface defined as
follows:

\begin{Shaded}
\begin{Highlighting}[]
\NormalTok{interface Packet \{}
\NormalTok{  sequence}\OperatorTok{:}\NormalTok{ uint64}
\NormalTok{  timeoutHeight}\OperatorTok{:}\NormalTok{ uint64}
\NormalTok{  timeoutTimestamp}\OperatorTok{:}\NormalTok{ uint64}
\NormalTok{  sourcePort}\OperatorTok{:}\NormalTok{ Identifier}
\NormalTok{  sourceChannel}\OperatorTok{:}\NormalTok{ Identifier}
\NormalTok{  destPort}\OperatorTok{:}\NormalTok{ Identifier}
\NormalTok{  destChannel}\OperatorTok{:}\NormalTok{ Identifier}
\NormalTok{  data}\OperatorTok{:}\NormalTok{ bytes}
\NormalTok{\}}
\end{Highlighting}
\end{Shaded}

\begin{itemize}
\tightlist
\item
  The \texttt{sequence} number corresponds to the order of sends and
  receives, where a packet with an earlier sequence number must be sent
  and received before a packet with a later sequence number.
\item
  The \texttt{timeoutHeight} indicates a consensus height on the
  destination ledger after which the packet will no longer be processed,
  and will instead count as having timed-out.
\item
  The \texttt{timeoutTimestamp} indicates a timestamp on the destination
  ledger after which the packet will no longer be processed, and will
  instead count as having timed-out.
\item
  The \texttt{sourcePort} identifies the port on the sending ledger.
\item
  The \texttt{sourceChannel} identifies the channel end on the sending
  ledger.
\item
  The \texttt{destPort} identifies the port on the receiving ledger.
\item
  The \texttt{destChannel} identifies the channel end on the receiving
  ledger.
\item
  The \texttt{data} is an opaque value which can be defined by the
  application logic of the associated modules.
\end{itemize}

Note that a \texttt{Packet} is never directly serialised. Rather it is
an intermediary structure used in certain function calls that may need
to be created or processed by modules calling the IBC handler.

\vspace{3mm}

\hypertarget{properties}{%
\subsubsection{Properties}\label{properties}}

\vspace{3mm}

\hypertarget{efficiency}{%
\paragraph{Efficiency}\label{efficiency}}

As channels impose no flow control of their own, the speed of packet
transmission and confirmation is limited only by the speed of the
underlying ledgers.

\vspace{3mm}

\hypertarget{exactly-once-delivery}{%
\paragraph{Exactly-once delivery}\label{exactly-once-delivery}}

IBC packets sent on one end of a channel are delivered no more than
exactly once to the other end. No network synchrony assumptions are
required for exactly-once safety. If one or both of the ledgers halt,
packets may be delivered no more than once, and once the ledgers resume
packets will be able to flow again.

\vspace{3mm}

\hypertarget{ordering}{%
\paragraph{Ordering}\label{ordering}}

On ordered channels, packets are be sent and received in the same order:
if packet \texttt{x} is sent before packet \texttt{y} by a channel end
on ledger A, packet \texttt{x} will be received before packet \texttt{y}
by the corresponding channel end on ledger B.

On unordered channels, packets may be sent and received in any order.
Unordered packets, like ordered packets, have individual timeouts
specified in terms of the destination ledger's height or timestamp.

\vspace{3mm}

\hypertarget{permissioning}{%
\paragraph{Permissioning}\label{permissioning}}

Channels are permissioned to one module on each end, determined during
the handshake and immutable afterwards (higher-level logic could
tokenise channel ownership by tokenising ownership of the port). Only
the module which owns the port associated with a channel end is able to
send or receive on the channel.

\vspace{3mm}

\hypertarget{channel-lifecycle-management}{%
\subsubsection{Channel lifecycle
management}\label{channel-lifecycle-management}}

\vspace{3mm}

\hypertarget{opening-handshake}{%
\paragraph{Opening handshake}\label{opening-handshake}}

The channel opening handshake, between two ledgers \texttt{A} and
\texttt{B}, with state formatted as \texttt{(A,\ B)}, flows as follows:

\begin{longtable}[]{@{}lll@{}}
\toprule
Datagram & Prior state & Posterior state\tabularnewline
\midrule
\endhead
\texttt{ChanOpenInit} & \texttt{(-,\ -)} &
\texttt{(INIT,\ -)}\tabularnewline
\texttt{ChanOpenTry} & \texttt{(INIT,\ -)} &
\texttt{(INIT,\ TRYOPEN)}\tabularnewline
\texttt{ChanOpenAck} & \texttt{(INIT,\ TRYOPEN)} &
\texttt{(OPEN,\ TRYOPEN)}\tabularnewline
\texttt{ChanOpenConfirm} & \texttt{(OPEN,\ TRYOPEN)} &
\texttt{(OPEN,\ OPEN)}\tabularnewline
\bottomrule
\end{longtable}

\texttt{ChanOpenInit}, executed on ledger A, initiates a channel opening
handshake from a module on ledger A to a module on ledger B, providing
the identifiers of the local channel identifier, local port, remote
port, and remote channel identifier. ledger A stores a channel end
object in its state.

\texttt{ChanOpenTry}, executed on ledger B, relays notice of a channel
handshake attempt to the module on ledger B, providing the pair of
channel identifiers, a pair of port identifiers, and a desired version.
ledger B verifies a proof that ledger A has stored these identifiers as
claimed, looks up the module which owns the destination port, calls that
module to check that the version requested is compatible, and stores a
channel end object in its state.

\texttt{ChanOpenAck}, executed on ledger A, relays acceptance of a
channel handshake attempt back to the module on ledger A, providing the
identifier which can now be used to look up the channel end. ledger A
verifies a proof that ledger B has stored the channel metadata as
claimed and marks its end of the channel as \texttt{OPEN}.

\texttt{ChanOpenConfirm}, executed on ledger B, confirms opening of a
channel from ledger A to ledger B. Ledger B simply checks that ledger A
has executed \texttt{ChanOpenAck} and marked the channel as
\texttt{OPEN}. Ledger B subsequently marks its end of the channel as
\texttt{OPEN}. After execution of \texttt{ChanOpenConfirm}, the channel
is open on both ends and can be used immediately.

When the opening handshake is complete, the module which initiates the
handshake will own the end of the created channel on the host ledger,
and the counterparty module which it specifies will own the other end of
the created channel on the counterparty ledger. Once a channel is
created, ownership can only be changed by changing ownership of the
associated ports.

\vspace{3mm}

\hypertarget{versioning}{%
\paragraph{Versioning}\label{versioning}}

During the handshake process, two ends of a channel come to agreement on
a version bytestring associated with that channel. The contents of this
version bytestring are opaque to the IBC core protocol. Host ledgers may
utilise the version data to indicate supported application-layer
protocols, agree on packet encoding formats, or negotiate other
channel-related metadata related to custom logic on top of IBC. Host
ledgers may also safely ignore the version data or specify an empty
string.

\vspace{3mm}

\hypertarget{closing-handshake}{%
\paragraph{Closing handshake}\label{closing-handshake}}

The channel closing handshake, between two ledgers \texttt{A} and
\texttt{B}, with state formatted as \texttt{(A,\ B)}, flows as follows:

\begin{longtable}[]{@{}lll@{}}
\toprule
Datagram & Prior state & Posterior state\tabularnewline
\midrule
\endhead
\texttt{ChanCloseInit} & \texttt{\small{(OPEN, OPEN)}} &
\texttt{\small{(CLOSED, OPEN)}}\tabularnewline
\texttt{ChanCloseConfirm} & \texttt{\small{(CLOSED, OPEN)}} &
\texttt{\small{(CLOSED, CLOSED)}}\tabularnewline
\bottomrule
\end{longtable}

\texttt{ChanCloseInit}, executed on ledger A, closes the end of the
channel on ledger A.

\texttt{ChanCloseInit}, executed on ledger B, simply verifies that the
channel has been marked as closed on ledger A and closes the end on
ledger B.

Any in-flight packets can be timed-out as soon as a channel is closed.

Once closed, channels cannot be reopened and identifiers cannot be
reused. Identifier reuse is prevented because we want to prevent
potential replay of previously sent packets. The replay problem is
analogous to using sequence numbers with signed messages, except where
the light client algorithm ``signs'' the messages (IBC packets), and the
replay prevention sequence is the combination of port identifier,
channel identifier, and packet sequence --- hence we cannot allow the
same port identifier and channel identifier to be reused again with a
sequence reset to zero, since this might allow packets to be replayed.
It would be possible to safely reuse identifiers if timeouts of a
particular maximum height/time were mandated and tracked, and future
protocol versions may incorporate this feature.

\vspace{3mm}

\hypertarget{sending-packets}{%
\subsubsection{Sending packets}\label{sending-packets}}

\vspace{3mm}

The \texttt{sendPacket} function is called by a module in order to send
an IBC packet on a channel end owned by the calling module to the
corresponding module on the counterparty ledger.

Calling modules must execute application logic atomically in conjunction
with calling \texttt{sendPacket}.

The IBC handler performs the following steps in order:

\begin{itemize}
\tightlist
\item
  Checks that the channel and connection are open to send packets
\item
  Checks that the calling module owns the sending port
\item
  Checks that the packet metadata matches the channel and connection
  information
\item
  Checks that the timeout height specified has not already passed on the
  destination ledger
\item
  Increments the send sequence counter associated with the channel (in
  the case of ordered channels)
\item
  Stores a constant-size commitment to the packet data and packet
  timeout
\end{itemize}

Note that the full packet is not stored in the state of the ledger ---
merely a short hash-commitment to the data and timeout value. The packet
data can be calculated from the transaction execution and possibly
returned as log output which relayers can index.

\vspace{3mm}

\hypertarget{receiving-packets}{%
\subsubsection{Receiving packets}\label{receiving-packets}}

\vspace{3mm}

The \texttt{recvPacket} function is called by a module in order to
receive and process an IBC packet sent on the corresponding channel end
on the counterparty ledger.

Calling modules must execute application logic atomically in conjunction
with calling \texttt{recvPacket}, likely beforehand to calculate the
acknowledgement value.

The IBC handler performs the following steps in order:

\begin{itemize}
\tightlist
\item
  Checks that the channel and connection are open to receive packets
\item
  Checks that the calling module owns the receiving port
\item
  Checks that the packet metadata matches the channel and connection
  information
\item
  Checks that the packet sequence is the next sequence the channel end
  expects to receive (for ordered channels)
\item
  Checks that the timeout height has not yet passed
\item
  Checks the inclusion proof of packet data commitment in the outgoing
  ledger's state
\item
  Sets the opaque acknowledgement value at a store path unique to the
  packet (if the acknowledgement is non-empty or the channel is
  unordered)
\item
  Increments the packet receive sequence associated with the channel end
  (for ordered channels)
\end{itemize}

\vspace{3mm}

\hypertarget{acknowledgements}{%
\paragraph{Acknowledgements}\label{acknowledgements}}

\vspace{3mm}

The \texttt{acknowledgePacket} function is called by a module to process
the acknowledgement of a packet previously sent by the calling module on
a channel to a counterparty module on the counterparty ledger.
\texttt{acknowledgePacket} also cleans up the packet commitment, which
is no longer necessary since the packet has been received and acted
upon.

Calling modules may atomically execute appropriate application
acknowledgement-handling logic in conjunction with calling
\texttt{acknowledgePacket}.

The IBC handler performs the following steps in order:

\begin{itemize}
\tightlist
\item
  Checks that the channel and connection are open to acknowledge packets
\item
  Checks that the calling module owns the sending port
\item
  Checks that the packet metadata matches the channel and connection
  information
\item
  Checks that the packet was actually sent on this channel
\item
  Checks that the packet sequence is the next sequence the channel end
  expects to acknowledge (for ordered channels)
\item
  Checks the inclusion proof of the packet acknowledgement data in the
  receiving ledger's state
\item
  Deletes the packet commitment (cleaning up state and preventing
  replay)
\item
  Increments the next acknowledgement sequence (for ordered channels)
\end{itemize}

\vspace{3mm}

\hypertarget{timeouts}{%
\subsubsection{Timeouts}\label{timeouts}}

\vspace{3mm}

Application semantics may require some timeout: an upper limit to how
long the ledger will wait for a transaction to be processed before
considering it an error. Since the two ledgers have different local
clocks, this is an obvious attack vector for a double spend --- an
attacker may delay the relay of the receipt or wait to send the packet
until right after the timeout --- so applications cannot safely
implement naive timeout logic themselves. In order to avoid any possible
``double-spend'' attacks, the timeout algorithm requires that the
destination ledger is running and reachable. The timeout must be proven
on the recipient ledger, not simply the absence of a response on the
sending ledger.

\vspace{3mm}

\hypertarget{sending-end}{%
\paragraph{Sending end}\label{sending-end}}

The \texttt{timeoutPacket} function is called by a module which
originally attempted to send a packet to a counterparty module, where
the timeout height or timeout timestamp has passed on the counterparty
ledger without the packet being committed, to prove that the packet can
no longer be executed and to allow the calling module to safely perform
appropriate state transitions.

Calling modules may atomically execute appropriate application
timeout-handling logic in conjunction with calling
\texttt{timeoutPacket}.

The IBC handler performs the following steps in order:

\begin{itemize}
\tightlist
\item
  Checks that the channel and connection are open to timeout packets
\item
  Checks that the calling module owns the sending port
\item
  Checks that the packet metadata matches the channel and connection
  information
\item
  Checks that the packet was actually sent on this channel
\item
  Checks a proof that the packet has not been confirmed on the
  destination ledger
\item
  Checks a proof that the destination ledger has exceeded the timeout
  height or timestamp
\item
  Deletes the packet commitment (cleaning up state and preventing
  replay)
\end{itemize}

In the case of an ordered channel, \texttt{timeoutPacket} additionally
closes the channel if a packet has timed out. Unordered channels are
expected to continue in the face of timed-out packets.

If relations are enforced between timeout heights of subsequent packets,
safe bulk timeouts of all packets prior to a timed-out packet can be
performed.

\vspace{3mm}

\hypertarget{timing-out-on-close}{%
\paragraph{Timing-out on close}\label{timing-out-on-close}}

If a channel is closed, in-flight packets can never be received and thus
can be safely timed-out. The \texttt{timeoutOnClose} function is called
by a module in order to prove that the channel to which an unreceived
packet was addressed has been closed, so the packet will never be
received (even if the \texttt{timeoutHeight} or
\texttt{timeoutTimestamp} has not yet been reached). Appropriate
application-specific logic may then safely be executed.

\vspace{3mm}

\hypertarget{cleaning-up-state}{%
\paragraph{Cleaning up state}\label{cleaning-up-state}}

If an acknowledgement is not written (as handling the acknowledgement
would clean up state in that case), \texttt{cleanupPacket} may be called
by a module in order to remove a received packet commitment from
storage. The receiving end must have already processed the packet
(whether regularly or past timeout).

In the ordered channel case, \texttt{cleanupPacket} cleans-up a packet
on an ordered channel by proving that the receive sequence has passed
the packet's sequence on the other end.

In the unordered channel case, \texttt{cleanupPacket} cleans-up a packet
on an unordered channel by proving that the associated acknowledgement
has been written.

\hypertarget{relayers}{%
\subsection{Relayers}\label{relayers}}

Relayer algorithms are the ``physical'' connection layer of IBC ---
off-ledger processes responsible for relaying data between two ledgers
running the IBC protocol by scanning the state of each ledger,
constructing appropriate datagrams, and executing them on the opposite
ledger as allowed by the protocol.

\vspace{3mm}

\hypertarget{motivation}{%
\subsubsection{Motivation}\label{motivation}}

\vspace{3mm}

In the IBC protocol, one ledger can only record the intention to send
particular data to another ledger --- it does not have direct access to
a network transport layer. Physical datagram relay must be performed by
off-ledger infrastructure with access to a transport layer such as
TCP/IP. This standard defines the concept of a \emph{relayer} algorithm,
executable by an off-ledger process with the ability to query ledger
state, to perform this relay.

A \emph{relayer} is an off-ledger process with the ability to read the
state of and submit transactions to some set of ledgers utilising the
IBC protocol.

\vspace{3mm}

\hypertarget{properties}{%
\subsubsection{Properties}\label{properties}}

\vspace{3mm}

\begin{itemize}
\tightlist
\item
  No exactly-once or deliver-or-timeout safety properties of IBC depend
  on relayer behaviour (Byzantine relayers are assumed)
\item
  Packet relay liveness properties of IBC depend only on the existence
  of at least one correct, live relayer
\item
  Relaying can safely be permissionless, all requisite verification is
  performed by the ledger itself
\item
  Requisite communication between the IBC user and the relayer is
  minimised
\item
  Provision for relayer incentivisation are not included in the core
  protocol, but are possible at the application layer
\end{itemize}

\vspace{3mm}

\hypertarget{basic-relayer-algorithm}{%
\subsubsection{Basic relayer algorithm}\label{basic-relayer-algorithm}}

\vspace{3mm}

The relayer algorithm is defined over a set of ledgers implementing the
IBC protocol. Each relayer may not necessarily have access to read state
from and write datagrams to all ledgers in the multi-ledger network
(especially in the case of permissioned or private ledgers) ---
different relayers may relay between different subsets.

Every so often, although no more frequently than once per block on
either ledger, a relayer calculates the set of all valid datagrams to be
relayed from one ledger to another based on the state of both ledgers.
The relayer must possess prior knowledge of what subset of the IBC
protocol is implemented by the ledgers in the set for which they are
relaying (e.g.~by reading the source code). Datagrams can be submitted
individually as single transactions or atomically as a single
transaction if the ledger supports it.

Different relayers may relay between different ledgers --- as long as
each pair of ledgers has at least one correct and live relayer and the
ledgers remain live, all packets flowing between ledgers in the network
will eventually be relayed.

\vspace{3mm}

\hypertarget{packets-acknowledgements-timeouts}{%
\subsubsection{Packets, acknowledgements,
timeouts}\label{packets-acknowledgements-timeouts}}

\vspace{3mm}

\hypertarget{relaying-packets-in-an-ordered-channel}{%
\paragraph{Relaying packets in an ordered
channel}\label{relaying-packets-in-an-ordered-channel}}

Packets in an ordered channel can be relayed in either an event-based
fashion or a query-based fashion. For the former, the relayer should
watch the source ledger for events emitted whenever packets are sent,
then compose the packet using the data in the event log. For the latter,
the relayer should periodically query the send sequence on the source
ledger, and keep the last sequence number relayed, so that any sequences
in between the two are packets that need to be queried and then relayed.
In either case, subsequently, the relayer process should check that the
destination ledger has not yet received the packet by checking the
receive sequence, and then relay it.

\vspace{3mm}

\hypertarget{relaying-packets-in-an-unordered-channel}{%
\paragraph{Relaying packets in an unordered
channel}\label{relaying-packets-in-an-unordered-channel}}

Packets in an unordered channel can most easily be relayed in an
event-based fashion. The relayer should watch the source ledger for
events emitted whenever packets are send, then compose the packet using
the data in the event log. Subsequently, the relayer should check
whether the destination ledger has received the packet already by
querying for the presence of an acknowledgement at the packet's sequence
number, and if one is not yet present the relayer should relay the
packet.

\vspace{3mm}

\hypertarget{relaying-acknowledgements}{%
\paragraph{Relaying acknowledgements}\label{relaying-acknowledgements}}

Acknowledgements can most easily be relayed in an event-based fashion.
The relayer should watch the destination ledger for events emitted
whenever packets are received and acknowledgements are written, then
compose the acknowledgement using the data in the event log, check
whether the packet commitment still exists on the source ledger (it will
be deleted once the acknowledgement is relayed), and if so relay the
acknowledgement to the source ledger.

\vspace{3mm}

\hypertarget{relaying-timeouts}{%
\paragraph{Relaying timeouts}\label{relaying-timeouts}}

Timeout relay is slightly more complex since there is no specific event
emitted when a packet times-out --- it is simply the case that the
packet can no longer be relayed, since the timeout height or timestamp
has passed on the destination ledger. The relayer process must elect to
track a set of packets (which can be constructed by scanning event
logs), and as soon as the height or timestamp of the destination ledger
exceeds that of a tracked packet, check whether the packet commitment
still exists on the source ledger (it will be deleted once the timeout
is relayed), and if so relay a timeout to the source ledger.

\vspace{3mm}

\hypertarget{ordering-constraints}{%
\paragraph{Ordering constraints}\label{ordering-constraints}}

There are implicit ordering constraints imposed on the relayer process
determining which datagrams must be submitted in what order. For
example, a header must be submitted to finalise the stored consensus
state and commitment root for a particular height in a light client
before a packet can be relayed. The relayer process is responsible for
frequently querying the state of the ledgers between which they are
relaying in order to determine what must be relayed when.

\vspace{3mm}

\hypertarget{bundling}{%
\paragraph{Bundling}\label{bundling}}

If the host ledger supports it, the relayer process can bundle many
datagrams into a single transaction, which will cause them to be
executed in sequence, and amortise any overhead costs (e.g.~signature
checks for fee payment).

\vspace{3mm}

\hypertarget{race-conditions}{%
\paragraph{Race conditions}\label{race-conditions}}

Multiple relayers relaying between the same pair of modules and ledgers
may attempt to relay the same packet (or submit the same header) at the
same time. If two relayers do so, the first transaction will succeed and
the second will fail. Out-of-band coordination between the relayers or
between the actors who sent the original packets and the relayers is
necessary to mitigate this.

\vspace{3mm}

\hypertarget{incentivisation}{%
\paragraph{Incentivisation}\label{incentivisation}}

The relay process must have access to accounts on both ledgers with
sufficient balance to pay for transaction fees. Relayers may employ
application-level methods to recoup these fees, such by including a
small payment to themselves in the packet data.

Any number of relayer processes may be safely run in parallel (and
indeed, it is expected that separate relayers will serve separate
subsets of the multi-ledger network). However, they may consume
unnecessary fees if they submit the same proof multiple times, so some
minimal coordination may be ideal (such as assigning particular relayers
to particular packets or scanning mempools for pending transactions).

\hypertarget{usage-patterns}{%
\section{Usage patterns}\label{usage-patterns}}

\hypertarget{call-receiver}{%
\subsection{Call receiver}\label{call-receiver}}

Essential to the functionality of the IBC handler is an interface to
other modules running on the same ledger, so that it can accept requests
to send packets and can route incoming packets to modules. This
interface should be as minimal as possible in order to reduce
implementation complexity and requirements imposed on host ledgers.

For this reason, the core IBC logic uses a receive-only call pattern
that differs slightly from the intuitive dataflow. As one might expect,
modules call into the IBC handler to create connections, channels, and
send packets. However, instead of the IBC handler, upon receipt of a
packet from another ledger, selecting and calling into the appropriate
module, the module itself must call \texttt{recvPacket} on the IBC
handler (likewise for accepting channel creation handshakes). When
\texttt{recvPacket} is called, the IBC handler will check that the
calling module is authorised to receive and process the packet (based on
included proofs and known state of connections / channels), perform
appropriate state updates (incrementing sequence numbers to prevent
replay), and return control to the module or throw on error. The IBC
handler never calls into modules directly.

Although a bit counterintuitive to reason about at first, this pattern
has a few notable advantages:

\begin{itemize}
\tightlist
\item
  It minimises requirements of the host ledger, since the IBC handler
  need not understand how to call into other modules or store any
  references to them.
\item
  It avoids the necessity of managing a module lookup table in the
  handler state.
\item
  It avoids the necessity of dealing with module return data or
  failures. If a module does not want to\\
  receive a packet (perhaps having implemented additional authorisation
  on top), it simply never calls \texttt{recvPacket}. If the routing
  logic were implemented in the IBC handler, the handler would need to
  deal with the failure of the module, which is tricky to interpret.
\end{itemize}

It also has one notable disadvantage: without an additional abstraction,
the relayer logic becomes more complex, since off-ledger relayer
processes will need to track the state of multiple modules to determine
when packets can be submitted.

For this reason, ledgers may implement an additional IBC ``routing
module'' which exposes a call dispatch interface.

\hypertarget{call-dispatch}{%
\subsection{Call dispatch}\label{call-dispatch}}

For common relay patterns, an ``IBC routing module'' can be implemented
which maintains a module dispatch table and simplifies the job of
relayers.

In the call dispatch pattern, datagrams (contained within transaction
types defined by the host ledger) are relayed directly to the routing
module, which then looks up the appropriate module (owning the channel
and port to which the datagram was addressed) and calls an appropriate
function (which must have been previously registered with the routing
module). This allows modules to avoid handling datagrams directly, and
makes it harder to accidentally screw-up the atomic state transition
execution which must happen in conjunction with sending or receiving a
packet (since the module never handles packets directly, but rather
exposes functions which are called by the routing module upon receipt of
a valid packet).

Additionally, the routing module can implement default logic for
handshake datagram handling (accepting incoming handshakes on behalf of
modules), which is convenient for modules which do not need to implement
their own custom logic.

\hypertarget{example-application-level-module}{%
\section{Example application-level
module}\label{example-application-level-module}}

The section specifies packet data structure and state machine handling
logic for the transfer of fungible tokens over an IBC channel between
two modules on separate ledgers. The state machine logic presented
allows for safe multi-ledger denomination handling with permissionless
channel opening. This logic constitutes a ``fungible token transfer
bridge module'', interfacing between the IBC routing module and an
existing asset tracking module on the host ledger.

\vspace{3mm}

\hypertarget{motivation}{%
\subsubsection{Motivation}\label{motivation}}

\vspace{3mm}

Users of a set of ledgers connected over the IBC protocol might wish to
utilise an asset issued on one ledger on another ledger, perhaps to make
use of additional features such as exchange or privacy protection, while
retaining fungibility with the original asset on the issuing ledger.
This application-layer protocol allows for transferring fungible tokens
between ledgers connected with IBC in a way which preserves asset
fungibility, preserves asset ownership, limits the impact of Byzantine
faults, and requires no additional permissioning.

\vspace{3mm}

\hypertarget{properties}{%
\subsubsection{Properties}\label{properties}}

\vspace{3mm}

\begin{itemize}
\tightlist
\item
  Preservation of fungibility (two-way peg)
\item
  Preservation of total supply (constant or inflationary on a single
  source ledger and module)
\item
  Permissionless token transfers, no need to whitelist connections,
  modules, or denominations
\item
  Symmetric (all ledgers implement the same logic)
\item
  Fault containment: prevents Byzantine-inflation of tokens originating
  on ledger A, as a result of ledger B's Byzantine behaviour (though any
  users who sent tokens to ledger B may be at risk)
\end{itemize}

\vspace{3mm}

\hypertarget{packet-definition}{%
\subsubsection{Packet definition}\label{packet-definition}}

\vspace{3mm}

Only one packet data type, \texttt{FungibleTokenPacketData}, which
specifies the denomination, amount, sending account, receiving account,
and whether the sending ledger is the source of the asset, is required:

\begin{Shaded}
\begin{Highlighting}[]
\NormalTok{interface FungibleTokenPacketData \{}
\NormalTok{  denomination}\OperatorTok{:} \DataTypeTok{string}
\NormalTok{  amount}\OperatorTok{:}\NormalTok{ uint256}
\NormalTok{  sender}\OperatorTok{:} \DataTypeTok{string}
\NormalTok{  receiver}\OperatorTok{:} \DataTypeTok{string}
\NormalTok{\}}
\end{Highlighting}
\end{Shaded}

The acknowledgement data type describes whether the transfer succeeded
or failed, and the reason for failure (if any):

\begin{Shaded}
\begin{Highlighting}[]
\NormalTok{interface FungibleTokenPacketAcknowledgement \{}
\NormalTok{  success}\OperatorTok{:} \DataTypeTok{boolean}
\NormalTok{  error}\OperatorTok{:}\NormalTok{ Maybe}\OperatorTok{\textless{}}\DataTypeTok{string}\OperatorTok{\textgreater{}}
\NormalTok{\}}
\end{Highlighting}
\end{Shaded}

\vspace{3mm}

\hypertarget{packet-handling-semantics}{%
\subsubsection{Packet handling
semantics}\label{packet-handling-semantics}}

\vspace{3mm}

The protocol logic is symmetric, so that denominations originating on
either ledger can be converted to vouchers on the other, and then
redeemed back again later.

\begin{itemize}
\tightlist
\item
  When acting as the source ledger, the bridge module escrows an
  existing local asset denomination on the sending ledger and mints
  vouchers on the receiving ledger.
\item
  When acting as the sink ledger, the bridge module burns local vouchers
  on the sending ledgers and unescrows the local asset denomination on
  the receiving ledger.
\item
  When a packet times-out, local assets are unescrowed back to the
  sender or vouchers minted back to the sender appropriately.
\item
  Acknowledgement data is used to handle failures, such as invalid
  denominations or invalid destination accounts. Returning an
  acknowledgement of failure is preferable to aborting the transaction
  since it more easily enables the sending ledger to take appropriate
  action based on the nature of the failure.
\end{itemize}

This implementation preserves both fungibility and supply. If tokens
have been sent to the counterparty ledger, they can be redeemed back in
the same denomination and amount on the source ledger. The combined
supply of unlocked tokens of a particular on both ledgers is constant,
since each send-receive packet pair locks and mints the same amount
(although the source ledger of a particular asset could change the
supply outside of the scope of this protocol).

\vspace{3mm}

\hypertarget{fault-containment}{%
\subsubsection{Fault containment}\label{fault-containment}}

\vspace{3mm}

Ledgers could fail to follow the fungible transfer token protocol
outlined here in one of two ways: the full nodes running the consensus
algorithm could diverge from the light client, or the ledger's state
machine could incorrectly implement the escrow \& voucher logic (whether
inadvertently or intentionally). Consensus divergence should eventually
result in evidence of misbehaviour which can be used to freeze the
client, but may not immediately do so (and no guarantee can be made that
such evidence would be submitted before more packets), so from the
perspective of the protocol's goal of isolating faults these cases must
be handled in the same way. No guarantees can be made about asset
recovery --- users electing to transfer tokens to a ledger take on the
risk of that ledger failing --- but containment logic can easily be
implemented on the interface boundary by tracking incoming and outgoing
supply of each asset, and ensuring that no ledger is allowed to redeem
vouchers for more tokens than it had initially escrowed. In essence,
particular channels can be treated as accounts, where a module on the
other end of a channel cannot spend more than it has received. Since
isolated Byzantine sub-graphs of a multi-ledger fungible token transfer
system will be unable to transfer out any more tokens than they had
initially received, this prevents any supply inflation of source assets,
and ensures that users only take on the consensus risk of ledgers they
intentionally connect to.

\vspace{3mm}

\hypertarget{multi-ledger-transfer-paths}{%
\subsubsection{Multi-ledger transfer
paths}\label{multi-ledger-transfer-paths}}

\vspace{3mm}

This protocol does not directly handle the ``diamond problem'', where a
user sends a token originating on ledger A to ledger B, then to ledger
D, and wants to return it through the path
\texttt{D\ -\textgreater{}\ C\ -\textgreater{}\ A} --- since the supply
is tracked as owned by ledger B (and the voucher denomination will be
\texttt{"\{portD\}/\{channelD\}/\{portB\}/\{channelB\}/denom"}), ledger
C cannot serve as the intermediary. This is necessary due to the fault
containment desiderata outlined above. Complexities arising from long
redemption paths may lead to the emergence of central ledgers in the
network topology or automated markets to exchange assets with different
redemption paths.

In order to track all of the denominations moving around the network of
ledgers in various paths, it may be helpful for a particular ledger to
implement a registry which will track the ``global'' source ledger for
each denomination. End-user service providers (such as wallet authors)
may want to integrate such a registry or keep their own mapping of
canonical source ledgers and human-readable names in order to improve
UX.

\hypertarget{testing-deployment}{%
\section{Testing \& deployment}\label{testing-deployment}}

A full version of the interblockchain protocol has been implemented in
Go in the Cosmos SDK {[}13{]}, an implementation is in progress in Rust
{[}14{]}, and implementations are planned for other languages in the
future. An off-ledger relayer daemon has also been implemented in Go
{[}15{]}. Game of Zones {[}16{]}, a live test of the initial software
release, is currently in progress. Over one hundred simulated zones
(separate consensus instances and ledgers) have been successfully linked
together {[}17{]}.

Production release and deployment to the Cosmos Network is planned for
later this summer. As IBC is a permissionless, opt-in protocol, adoption
will be dependent on ledgers voluntarily electing to support the
specification, in full or in part. Adoption of IBC does not require
connection to the Cosmos Hub, usage of any particular token, or even
usage of any other piece of Cosmos software --- IBC can be implemented
on top of other state machine frameworks such as Substrate {[}18{]}, or
by standalone ledgers using custom logic --- adherence to the correct
protocol is both necessary and sufficient for successful interoperation.

\hypertarget{acknowledgements}{%
\section{Acknowledgements}\label{acknowledgements}}

The original idea of IBC was first outlined in the Cosmos whitepaper
{[}19{]}, and realisation of the protocol is made possible in practice
by the Byzantine-fault-tolerant consensus and efficient light client
verification of Tendermint, introduced in \emph{Tendermint: Consensus
without Mining} {[}8{]} and updated in \emph{The latest gossip on BFT
consensus} {[}20{]}. An earlier version of the IBC specification
{[}21{]} was written by Ethan Frey.

Many current and former employees of All in Bits (dba Tendermint Inc.),
Agoric Systems, the Interchain Foundation, Informal Systems, and
Interchain GmbH participated in brainstorming and reviews of the IBC
protocol. Thanks are due in particular to Ethan Buchman, Jae Kwon, Ethan
Frey, Juwoon Yun, Anca Zamfir, Zarko Milosevic, Zaki Manian, Aditya
Sripal, Federico Kunze, Dean Tribble, Mark Miller, Brian Warner, Chris
Hibbert, Michael FIG, Sunny Aggarwal, Dev Ojha, Colin Axner, and Jack
Zampolin. Thanks also to Meher Roy at Chorus One. Thanks to Zaki Manian,
Sam Hart, and Adi Seredinschi for reviewing this paper.

This work was supported by the Interchain Foundation.

\onecolumn

\hypertarget{appendices}{%
\section{Appendices}\label{appendices}}

\hypertarget{connection-handshake}{%
\subsection{Connection handshake}\label{connection-handshake}}

\hypertarget{initiating-a-handshake}{%
\subsubsection{Initiating a handshake}\label{initiating-a-handshake}}

\vspace{3mm}

\begin{Shaded}
\begin{Highlighting}[]
\KeywordTok{function} \FunctionTok{connOpenInit}\NormalTok{(}
\NormalTok{  identifier}\OperatorTok{:}\NormalTok{ Identifier}\OperatorTok{,}
\NormalTok{  desiredCounterpartyConnectionIdentifier}\OperatorTok{:}\NormalTok{ Identifier}\OperatorTok{,}
\NormalTok{  counterpartyPrefix}\OperatorTok{:}\NormalTok{ CommitmentPrefix}\OperatorTok{,}
\NormalTok{  clientIdentifier}\OperatorTok{:}\NormalTok{ Identifier}\OperatorTok{,}
\NormalTok{  counterpartyClientIdentifier}\OperatorTok{:}\NormalTok{ Identifier) \{}
    \FunctionTok{abortTransactionUnless}\NormalTok{(}\FunctionTok{validateConnectionIdentifier}\NormalTok{(identifier))}
    \FunctionTok{abortTransactionUnless}\NormalTok{(provableStore}\OperatorTok{.}\FunctionTok{get}\NormalTok{(}\FunctionTok{connectionPath}\NormalTok{(identifier)) }\OperatorTok{==}\NormalTok{ null)}
\NormalTok{    state }\OperatorTok{=}\NormalTok{ INIT}
\NormalTok{    connection }\OperatorTok{=}\NormalTok{ ConnectionEnd\{state}\OperatorTok{,}\NormalTok{ desiredCounterpartyConnectionIdentifier}\OperatorTok{,}\NormalTok{ counterpartyPrefix}\OperatorTok{,}
\NormalTok{      clientIdentifier}\OperatorTok{,}\NormalTok{ counterpartyClientIdentifier}\OperatorTok{,} \FunctionTok{getCompatibleVersions}\NormalTok{()\}}
\NormalTok{    provableStore}\OperatorTok{.}\FunctionTok{set}\NormalTok{(}\FunctionTok{connectionPath}\NormalTok{(identifier)}\OperatorTok{,}\NormalTok{ connection)}
\NormalTok{\}}
\end{Highlighting}
\end{Shaded}

\hypertarget{responding-to-a-handshake-initiation}{%
\subsubsection{Responding to a handshake
initiation}\label{responding-to-a-handshake-initiation}}

\vspace{3mm}

\begin{Shaded}
\begin{Highlighting}[]
\KeywordTok{function} \FunctionTok{connOpenTry}\NormalTok{(}
\NormalTok{  desiredIdentifier}\OperatorTok{:}\NormalTok{ Identifier}\OperatorTok{,}
\NormalTok{  counterpartyConnectionIdentifier}\OperatorTok{:}\NormalTok{ Identifier}\OperatorTok{,}
\NormalTok{  counterpartyPrefix}\OperatorTok{:}\NormalTok{ CommitmentPrefix}\OperatorTok{,}
\NormalTok{  counterpartyClientIdentifier}\OperatorTok{:}\NormalTok{ Identifier}\OperatorTok{,}
\NormalTok{  clientIdentifier}\OperatorTok{:}\NormalTok{ Identifier}\OperatorTok{,}
\NormalTok{  counterpartyVersions}\OperatorTok{:} \DataTypeTok{string}\NormalTok{[]}\OperatorTok{,}
\NormalTok{  proofInit}\OperatorTok{:}\NormalTok{ CommitmentProof}\OperatorTok{,}
\NormalTok{  proofConsensus}\OperatorTok{:}\NormalTok{ CommitmentProof}\OperatorTok{,}
\NormalTok{  proofHeight}\OperatorTok{:}\NormalTok{ uint64}\OperatorTok{,}
\NormalTok{  consensusHeight}\OperatorTok{:}\NormalTok{ uint64) \{}
    \FunctionTok{abortTransactionUnless}\NormalTok{(}\FunctionTok{validateConnectionIdentifier}\NormalTok{(desiredIdentifier))}
    \FunctionTok{abortTransactionUnless}\NormalTok{(consensusHeight }\OperatorTok{\textless{}=} \FunctionTok{getCurrentHeight}\NormalTok{())}
\NormalTok{    expectedConsensusState }\OperatorTok{=} \FunctionTok{getConsensusState}\NormalTok{(consensusHeight)}
\NormalTok{    expected }\OperatorTok{=}\NormalTok{ ConnectionEnd\{INIT}\OperatorTok{,}\NormalTok{ desiredIdentifier}\OperatorTok{,} \FunctionTok{getCommitmentPrefix}\NormalTok{()}\OperatorTok{,}\NormalTok{ counterpartyClientIdentifier}\OperatorTok{,}
\NormalTok{                             clientIdentifier}\OperatorTok{,}\NormalTok{ counterpartyVersions\}}
\NormalTok{    version }\OperatorTok{=} \FunctionTok{pickVersion}\NormalTok{(counterpartyVersions)}
\NormalTok{    connection }\OperatorTok{=}\NormalTok{ ConnectionEnd\{TRYOPEN}\OperatorTok{,}\NormalTok{ counterpartyConnectionIdentifier}\OperatorTok{,}\NormalTok{ counterpartyPrefix}\OperatorTok{,}
\NormalTok{                               clientIdentifier}\OperatorTok{,}\NormalTok{ counterpartyClientIdentifier}\OperatorTok{,}\NormalTok{ version\}}
    \FunctionTok{abortTransactionUnless}\NormalTok{(}
\NormalTok{      connection}\OperatorTok{.}\FunctionTok{verifyConnectionState}\NormalTok{(proofHeight}\OperatorTok{,}\NormalTok{ proofInit}\OperatorTok{,}\NormalTok{ counterpartyConnectionIdentifier}\OperatorTok{,}\NormalTok{ expected))}
    \FunctionTok{abortTransactionUnless}\NormalTok{(connection}\OperatorTok{.}\FunctionTok{verifyClientConsensusState}\NormalTok{(}
\NormalTok{      proofHeight}\OperatorTok{,}\NormalTok{ proofConsensus}\OperatorTok{,}\NormalTok{ counterpartyClientIdentifier}\OperatorTok{,}\NormalTok{ consensusHeight}\OperatorTok{,}\NormalTok{ expectedConsensusState))}
\NormalTok{    previous }\OperatorTok{=}\NormalTok{ provableStore}\OperatorTok{.}\FunctionTok{get}\NormalTok{(}\FunctionTok{connectionPath}\NormalTok{(desiredIdentifier))}
    \FunctionTok{abortTransactionUnless}\NormalTok{(}
\NormalTok{      (previous }\OperatorTok{===}\NormalTok{ null) }\OperatorTok{||}
\NormalTok{      (previous}\OperatorTok{.}\AttributeTok{state} \OperatorTok{===}\NormalTok{ INIT }\OperatorTok{\&\&}
\NormalTok{        previous}\OperatorTok{.}\AttributeTok{counterpartyConnectionIdentifier} \OperatorTok{===}\NormalTok{ counterpartyConnectionIdentifier }\OperatorTok{\&\&}
\NormalTok{        previous}\OperatorTok{.}\AttributeTok{counterpartyPrefix} \OperatorTok{===}\NormalTok{ counterpartyPrefix }\OperatorTok{\&\&}
\NormalTok{        previous}\OperatorTok{.}\AttributeTok{clientIdentifier} \OperatorTok{===}\NormalTok{ clientIdentifier }\OperatorTok{\&\&}
\NormalTok{        previous}\OperatorTok{.}\AttributeTok{counterpartyClientIdentifier} \OperatorTok{===}\NormalTok{ counterpartyClientIdentifier }\OperatorTok{\&\&}
\NormalTok{        previous}\OperatorTok{.}\AttributeTok{version} \OperatorTok{===}\NormalTok{ version))}
\NormalTok{    identifier }\OperatorTok{=}\NormalTok{ desiredIdentifier}
\NormalTok{    provableStore}\OperatorTok{.}\FunctionTok{set}\NormalTok{(}\FunctionTok{connectionPath}\NormalTok{(identifier)}\OperatorTok{,}\NormalTok{ connection)}
\NormalTok{\}}
\end{Highlighting}
\end{Shaded}

\hypertarget{acknowledging-the-response}{%
\subsubsection{Acknowledging the
response}\label{acknowledging-the-response}}

\vspace{3mm}

\begin{Shaded}
\begin{Highlighting}[]
\KeywordTok{function} \FunctionTok{connOpenAck}\NormalTok{(}
\NormalTok{  identifier}\OperatorTok{:}\NormalTok{ Identifier}\OperatorTok{,}
\NormalTok{  version}\OperatorTok{:} \DataTypeTok{string}\OperatorTok{,}
\NormalTok{  proofTry}\OperatorTok{:}\NormalTok{ CommitmentProof}\OperatorTok{,}    
\NormalTok{  proofConsensus}\OperatorTok{:}\NormalTok{ CommitmentProof}\OperatorTok{,}
\NormalTok{  proofHeight}\OperatorTok{:}\NormalTok{ uint64}\OperatorTok{,}
\NormalTok{  consensusHeight}\OperatorTok{:}\NormalTok{ uint64) \{    }
    \FunctionTok{abortTransactionUnless}\NormalTok{(consensusHeight }\OperatorTok{\textless{}=} \FunctionTok{getCurrentHeight}\NormalTok{())}
\NormalTok{    connection }\OperatorTok{=}\NormalTok{ provableStore}\OperatorTok{.}\FunctionTok{get}\NormalTok{(}\FunctionTok{connectionPath}\NormalTok{(identifier))}
    \FunctionTok{abortTransactionUnless}\NormalTok{(connection}\OperatorTok{.}\AttributeTok{state} \OperatorTok{===}\NormalTok{ INIT }\OperatorTok{||}\NormalTok{ connection}\OperatorTok{.}\AttributeTok{state} \OperatorTok{===}\NormalTok{ TRYOPEN)}
\NormalTok{    expectedConsensusState }\OperatorTok{=} \FunctionTok{getConsensusState}\NormalTok{(consensusHeight)}
\NormalTok{    expected }\OperatorTok{=}\NormalTok{ ConnectionEnd\{TRYOPEN}\OperatorTok{,}\NormalTok{ identifier}\OperatorTok{,} \FunctionTok{getCommitmentPrefix}\NormalTok{()}\OperatorTok{,}
\NormalTok{                             connection}\OperatorTok{.}\AttributeTok{counterpartyClientIdentifier}\OperatorTok{,}\NormalTok{ connection}\OperatorTok{.}\AttributeTok{clientIdentifier}\OperatorTok{,}
\NormalTok{                             version\} }
    \FunctionTok{abortTransactionUnless}\NormalTok{(connection}\OperatorTok{.}\FunctionTok{verifyConnectionState}\NormalTok{(proofHeight}\OperatorTok{,}\NormalTok{ proofTry}\OperatorTok{,}
\NormalTok{      connection}\OperatorTok{.}\AttributeTok{counterpartyConnectionIdentifier}\OperatorTok{,}\NormalTok{ expected))}
    \FunctionTok{abortTransactionUnless}\NormalTok{(connection}\OperatorTok{.}\FunctionTok{verifyClientConsensusState}\NormalTok{(}
\NormalTok{      proofHeight}\OperatorTok{,}\NormalTok{ proofConsensus}\OperatorTok{,}\NormalTok{ connection}\OperatorTok{.}\AttributeTok{counterpartyClientIdentifier}\OperatorTok{,}
\NormalTok{      consensusHeight}\OperatorTok{,}\NormalTok{ expectedConsensusState))}
\NormalTok{    connection}\OperatorTok{.}\AttributeTok{state} \OperatorTok{=}\NormalTok{ OPEN}
    \FunctionTok{abortTransactionUnless}\NormalTok{(}\FunctionTok{getCompatibleVersions}\NormalTok{()}\OperatorTok{.}\FunctionTok{indexOf}\NormalTok{(version) }\OperatorTok{!==} \OperatorTok{{-}}\DecValTok{1}\NormalTok{)}
\NormalTok{    connection}\OperatorTok{.}\AttributeTok{version} \OperatorTok{=}\NormalTok{ version}
\NormalTok{    provableStore}\OperatorTok{.}\FunctionTok{set}\NormalTok{(}\FunctionTok{connectionPath}\NormalTok{(identifier)}\OperatorTok{,}\NormalTok{ connection)}
\NormalTok{\}}
\end{Highlighting}
\end{Shaded}

\hypertarget{finalising-the-connection}{%
\subsubsection{Finalising the
connection}\label{finalising-the-connection}}

\vspace{3mm}

\begin{Shaded}
\begin{Highlighting}[]
\KeywordTok{function} \FunctionTok{connOpenConfirm}\NormalTok{(}
\NormalTok{  identifier}\OperatorTok{:}\NormalTok{ Identifier}\OperatorTok{,}
\NormalTok{  proofAck}\OperatorTok{:}\NormalTok{ CommitmentProof}\OperatorTok{,}
\NormalTok{  proofHeight}\OperatorTok{:}\NormalTok{ uint64) \{}
\NormalTok{    connection }\OperatorTok{=}\NormalTok{ provableStore}\OperatorTok{.}\FunctionTok{get}\NormalTok{(}\FunctionTok{connectionPath}\NormalTok{(identifier))}
    \FunctionTok{abortTransactionUnless}\NormalTok{(connection}\OperatorTok{.}\AttributeTok{state} \OperatorTok{===}\NormalTok{ TRYOPEN)}
\NormalTok{    expected }\OperatorTok{=}\NormalTok{ ConnectionEnd\{OPEN}\OperatorTok{,}\NormalTok{ identifier}\OperatorTok{,} \FunctionTok{getCommitmentPrefix}\NormalTok{()}\OperatorTok{,}
\NormalTok{                             connection}\OperatorTok{.}\AttributeTok{counterpartyClientIdentifier}\OperatorTok{,}
\NormalTok{                             connection}\OperatorTok{.}\AttributeTok{clientIdentifier}\OperatorTok{,}\NormalTok{ connection}\OperatorTok{.}\AttributeTok{version}\NormalTok{\}}
    \FunctionTok{abortTransactionUnless}\NormalTok{(connection}\OperatorTok{.}\FunctionTok{verifyConnectionState}\NormalTok{(}
\NormalTok{      proofHeight}\OperatorTok{,}\NormalTok{ proofAck}\OperatorTok{,}\NormalTok{ connection}\OperatorTok{.}\AttributeTok{counterpartyConnectionIdentifier}\OperatorTok{,}\NormalTok{ expected))}
\NormalTok{    connection}\OperatorTok{.}\AttributeTok{state} \OperatorTok{=}\NormalTok{ OPEN}
\NormalTok{    provableStore}\OperatorTok{.}\FunctionTok{set}\NormalTok{(}\FunctionTok{connectionPath}\NormalTok{(identifier)}\OperatorTok{,}\NormalTok{ connection)}
\NormalTok{\}}
\end{Highlighting}
\end{Shaded}

\hypertarget{channel-handshake}{%
\subsection{Channel handshake}\label{channel-handshake}}

\hypertarget{initiating-a-handshake}{%
\subsubsection{Initiating a handshake}\label{initiating-a-handshake}}

\vspace{3mm}

\begin{Shaded}
\begin{Highlighting}[]
\KeywordTok{function} \FunctionTok{chanOpenInit}\NormalTok{(}
\NormalTok{  order}\OperatorTok{:}\NormalTok{ ChannelOrder}\OperatorTok{,}
\NormalTok{  connectionHops}\OperatorTok{:}\NormalTok{ [Identifier]}\OperatorTok{,}
\NormalTok{  portIdentifier}\OperatorTok{:}\NormalTok{ Identifier}\OperatorTok{,}
\NormalTok{  channelIdentifier}\OperatorTok{:}\NormalTok{ Identifier}\OperatorTok{,}
\NormalTok{  counterpartyPortIdentifier}\OperatorTok{:}\NormalTok{ Identifier}\OperatorTok{,}
\NormalTok{  counterpartyChannelIdentifier}\OperatorTok{:}\NormalTok{ Identifier}\OperatorTok{,}
\NormalTok{  version}\OperatorTok{:} \DataTypeTok{string}\NormalTok{)}\OperatorTok{:}\NormalTok{ CapabilityKey \{}
    \FunctionTok{abortTransactionUnless}\NormalTok{(}\FunctionTok{validateChannelIdentifier}\NormalTok{(portIdentifier}\OperatorTok{,}\NormalTok{ channelIdentifier))}
    \FunctionTok{abortTransactionUnless}\NormalTok{(connectionHops}\OperatorTok{.}\AttributeTok{length} \OperatorTok{===} \DecValTok{1}\NormalTok{)}
    \FunctionTok{abortTransactionUnless}\NormalTok{(provableStore}\OperatorTok{.}\FunctionTok{get}\NormalTok{(}\FunctionTok{channelPath}\NormalTok{(portIdentifier}\OperatorTok{,}\NormalTok{ channelIdentifier)) }\OperatorTok{===}\NormalTok{ null)}
\NormalTok{    connection }\OperatorTok{=}\NormalTok{ provableStore}\OperatorTok{.}\FunctionTok{get}\NormalTok{(}\FunctionTok{connectionPath}\NormalTok{(connectionHops[}\DecValTok{0}\NormalTok{]))}
    \FunctionTok{abortTransactionUnless}\NormalTok{(connection }\OperatorTok{!==}\NormalTok{ null)}
    \FunctionTok{abortTransactionUnless}\NormalTok{(}\FunctionTok{authenticateCapability}\NormalTok{(}\FunctionTok{portPath}\NormalTok{(portIdentifier)}\OperatorTok{,}\NormalTok{ portCapability))}
\NormalTok{    channel }\OperatorTok{=}\NormalTok{ ChannelEnd\{INIT}\OperatorTok{,}\NormalTok{ order}\OperatorTok{,}\NormalTok{ counterpartyPortIdentifier}\OperatorTok{,}
\NormalTok{                         counterpartyChannelIdentifier}\OperatorTok{,}\NormalTok{ connectionHops}\OperatorTok{,}\NormalTok{ version\}}
\NormalTok{    provableStore}\OperatorTok{.}\FunctionTok{set}\NormalTok{(}\FunctionTok{channelPath}\NormalTok{(portIdentifier}\OperatorTok{,}\NormalTok{ channelIdentifier)}\OperatorTok{,}\NormalTok{ channel)}
\NormalTok{    channelCapability }\OperatorTok{=} \FunctionTok{newCapability}\NormalTok{(}\FunctionTok{channelCapabilityPath}\NormalTok{(portIdentifier}\OperatorTok{,}\NormalTok{ channelIdentifier))}
\NormalTok{    provableStore}\OperatorTok{.}\FunctionTok{set}\NormalTok{(}\FunctionTok{nextSequenceSendPath}\NormalTok{(portIdentifier}\OperatorTok{,}\NormalTok{ channelIdentifier)}\OperatorTok{,} \DecValTok{1}\NormalTok{)}
\NormalTok{    provableStore}\OperatorTok{.}\FunctionTok{set}\NormalTok{(}\FunctionTok{nextSequenceRecvPath}\NormalTok{(portIdentifier}\OperatorTok{,}\NormalTok{ channelIdentifier)}\OperatorTok{,} \DecValTok{1}\NormalTok{)}
\NormalTok{    provableStore}\OperatorTok{.}\FunctionTok{set}\NormalTok{(}\FunctionTok{nextSequenceAckPath}\NormalTok{(portIdentifier}\OperatorTok{,}\NormalTok{ channelIdentifier)}\OperatorTok{,} \DecValTok{1}\NormalTok{)}
\NormalTok{    return channelCapability}
\NormalTok{\}}
\end{Highlighting}
\end{Shaded}

\hypertarget{responding-to-a-handshake-initiation}{%
\subsubsection{Responding to a handshake
initiation}\label{responding-to-a-handshake-initiation}}

\vspace{3mm}

\begin{Shaded}
\begin{Highlighting}[]
\KeywordTok{function} \FunctionTok{chanOpenTry}\NormalTok{(}
\NormalTok{  order}\OperatorTok{:}\NormalTok{ ChannelOrder}\OperatorTok{,}
\NormalTok{  connectionHops}\OperatorTok{:}\NormalTok{ [Identifier]}\OperatorTok{,}
\NormalTok{  portIdentifier}\OperatorTok{:}\NormalTok{ Identifier}\OperatorTok{,}
\NormalTok{  channelIdentifier}\OperatorTok{:}\NormalTok{ Identifier}\OperatorTok{,}
\NormalTok{  counterpartyPortIdentifier}\OperatorTok{:}\NormalTok{ Identifier}\OperatorTok{,}
\NormalTok{  counterpartyChannelIdentifier}\OperatorTok{:}\NormalTok{ Identifier}\OperatorTok{,}
\NormalTok{  version}\OperatorTok{:} \DataTypeTok{string}\OperatorTok{,}
\NormalTok{  counterpartyVersion}\OperatorTok{:} \DataTypeTok{string}\OperatorTok{,}
\NormalTok{  proofInit}\OperatorTok{:}\NormalTok{ CommitmentProof}\OperatorTok{,}
\NormalTok{  proofHeight}\OperatorTok{:}\NormalTok{ uint64)}\OperatorTok{:}\NormalTok{ CapabilityKey \{}
    \FunctionTok{abortTransactionUnless}\NormalTok{(}\FunctionTok{validateChannelIdentifier}\NormalTok{(portIdentifier}\OperatorTok{,}\NormalTok{ channelIdentifier))}
    \FunctionTok{abortTransactionUnless}\NormalTok{(connectionHops}\OperatorTok{.}\AttributeTok{length} \OperatorTok{===} \DecValTok{1}\NormalTok{)}
\NormalTok{    previous }\OperatorTok{=}\NormalTok{ provableStore}\OperatorTok{.}\FunctionTok{get}\NormalTok{(}\FunctionTok{channelPath}\NormalTok{(portIdentifier}\OperatorTok{,}\NormalTok{ channelIdentifier))}
    \FunctionTok{abortTransactionUnless}\NormalTok{(}
\NormalTok{      (previous }\OperatorTok{===}\NormalTok{ null) }\OperatorTok{||}
\NormalTok{      (previous}\OperatorTok{.}\AttributeTok{state} \OperatorTok{===}\NormalTok{ INIT }\OperatorTok{\&\&}
\NormalTok{       previous}\OperatorTok{.}\AttributeTok{order} \OperatorTok{===}\NormalTok{ order }\OperatorTok{\&\&}
\NormalTok{       previous}\OperatorTok{.}\AttributeTok{counterpartyPortIdentifier} \OperatorTok{===}\NormalTok{ counterpartyPortIdentifier }\OperatorTok{\&\&}
\NormalTok{       previous}\OperatorTok{.}\AttributeTok{counterpartyChannelIdentifier} \OperatorTok{===}\NormalTok{ counterpartyChannelIdentifier }\OperatorTok{\&\&}
\NormalTok{       previous}\OperatorTok{.}\AttributeTok{connectionHops} \OperatorTok{===}\NormalTok{ connectionHops }\OperatorTok{\&\&}
\NormalTok{       previous}\OperatorTok{.}\AttributeTok{version} \OperatorTok{===}\NormalTok{ version)}
\NormalTok{      )}
    \FunctionTok{abortTransactionUnless}\NormalTok{(}\FunctionTok{authenticateCapability}\NormalTok{(}\FunctionTok{portPath}\NormalTok{(portIdentifier)}\OperatorTok{,}\NormalTok{ portCapability))}
\NormalTok{    connection }\OperatorTok{=}\NormalTok{ provableStore}\OperatorTok{.}\FunctionTok{get}\NormalTok{(}\FunctionTok{connectionPath}\NormalTok{(connectionHops[}\DecValTok{0}\NormalTok{]))}
    \FunctionTok{abortTransactionUnless}\NormalTok{(connection }\OperatorTok{!==}\NormalTok{ null)}
    \FunctionTok{abortTransactionUnless}\NormalTok{(connection}\OperatorTok{.}\AttributeTok{state} \OperatorTok{===}\NormalTok{ OPEN)}
\NormalTok{    expected }\OperatorTok{=}\NormalTok{ ChannelEnd\{INIT}\OperatorTok{,}\NormalTok{ order}\OperatorTok{,}\NormalTok{ portIdentifier}\OperatorTok{,}
\NormalTok{                          channelIdentifier}\OperatorTok{,}
\NormalTok{                          [connection}\OperatorTok{.}\AttributeTok{counterpartyConnectionIdentifier}\NormalTok{]}\OperatorTok{,}
\NormalTok{                          counterpartyVersion\}}
    \FunctionTok{abortTransactionUnless}\NormalTok{(connection}\OperatorTok{.}\FunctionTok{verifyChannelState}\NormalTok{(}
\NormalTok{      proofHeight}\OperatorTok{,}
\NormalTok{      proofInit}\OperatorTok{,}
\NormalTok{      counterpartyPortIdentifier}\OperatorTok{,}
\NormalTok{      counterpartyChannelIdentifier}\OperatorTok{,}
\NormalTok{      expected}
\NormalTok{    ))}
\NormalTok{    channel }\OperatorTok{=}\NormalTok{ ChannelEnd\{TRYOPEN}\OperatorTok{,}\NormalTok{ order}\OperatorTok{,}\NormalTok{ counterpartyPortIdentifier}\OperatorTok{,}
\NormalTok{                         counterpartyChannelIdentifier}\OperatorTok{,}\NormalTok{ connectionHops}\OperatorTok{,}\NormalTok{ version\}}
\NormalTok{    provableStore}\OperatorTok{.}\FunctionTok{set}\NormalTok{(}\FunctionTok{channelPath}\NormalTok{(portIdentifier}\OperatorTok{,}\NormalTok{ channelIdentifier)}\OperatorTok{,}\NormalTok{ channel)}
\NormalTok{    channelCapability }\OperatorTok{=} \FunctionTok{newCapability}\NormalTok{(}\FunctionTok{channelCapabilityPath}\NormalTok{(portIdentifier}\OperatorTok{,}\NormalTok{ channelIdentifier))}
\NormalTok{    provableStore}\OperatorTok{.}\FunctionTok{set}\NormalTok{(}\FunctionTok{nextSequenceSendPath}\NormalTok{(portIdentifier}\OperatorTok{,}\NormalTok{ channelIdentifier)}\OperatorTok{,} \DecValTok{1}\NormalTok{)}
\NormalTok{    provableStore}\OperatorTok{.}\FunctionTok{set}\NormalTok{(}\FunctionTok{nextSequenceRecvPath}\NormalTok{(portIdentifier}\OperatorTok{,}\NormalTok{ channelIdentifier)}\OperatorTok{,} \DecValTok{1}\NormalTok{)}
\NormalTok{    provableStore}\OperatorTok{.}\FunctionTok{set}\NormalTok{(}\FunctionTok{nextSequenceAckPath}\NormalTok{(portIdentifier}\OperatorTok{,}\NormalTok{ channelIdentifier)}\OperatorTok{,} \DecValTok{1}\NormalTok{)}
\NormalTok{    return channelCapability}
\NormalTok{\}}
\end{Highlighting}
\end{Shaded}

\hypertarget{acknowledging-the-response}{%
\subsubsection{Acknowledging the
response}\label{acknowledging-the-response}}

\vspace{3mm}

\begin{Shaded}
\begin{Highlighting}[]
\KeywordTok{function} \FunctionTok{chanOpenAck}\NormalTok{(}
\NormalTok{  portIdentifier}\OperatorTok{:}\NormalTok{ Identifier}\OperatorTok{,}
\NormalTok{  channelIdentifier}\OperatorTok{:}\NormalTok{ Identifier}\OperatorTok{,}
\NormalTok{  counterpartyVersion}\OperatorTok{:} \DataTypeTok{string}\OperatorTok{,}
\NormalTok{  proofTry}\OperatorTok{:}\NormalTok{ CommitmentProof}\OperatorTok{,}
\NormalTok{  proofHeight}\OperatorTok{:}\NormalTok{ uint64) \{}
\NormalTok{    channel }\OperatorTok{=}\NormalTok{ provableStore}\OperatorTok{.}\FunctionTok{get}\NormalTok{(}\FunctionTok{channelPath}\NormalTok{(portIdentifier}\OperatorTok{,}\NormalTok{ channelIdentifier))}
    \FunctionTok{abortTransactionUnless}\NormalTok{(channel}\OperatorTok{.}\AttributeTok{state} \OperatorTok{===}\NormalTok{ INIT }\OperatorTok{||}\NormalTok{ channel}\OperatorTok{.}\AttributeTok{state} \OperatorTok{===}\NormalTok{ TRYOPEN)}
    \FunctionTok{abortTransactionUnless}\NormalTok{(}\FunctionTok{authenticateCapability}\NormalTok{(}\FunctionTok{channelCapabilityPath}\NormalTok{(portIdentifier}\OperatorTok{,}\NormalTok{ channelIdentifier)}\OperatorTok{,}\NormalTok{ capability))}
\NormalTok{    connection }\OperatorTok{=}\NormalTok{ provableStore}\OperatorTok{.}\FunctionTok{get}\NormalTok{(}\FunctionTok{connectionPath}\NormalTok{(channel}\OperatorTok{.}\AttributeTok{connectionHops}\NormalTok{[}\DecValTok{0}\NormalTok{]))}
    \FunctionTok{abortTransactionUnless}\NormalTok{(connection }\OperatorTok{!==}\NormalTok{ null)}
    \FunctionTok{abortTransactionUnless}\NormalTok{(connection}\OperatorTok{.}\AttributeTok{state} \OperatorTok{===}\NormalTok{ OPEN)}
\NormalTok{    expected }\OperatorTok{=}\NormalTok{ ChannelEnd\{TRYOPEN}\OperatorTok{,}\NormalTok{ channel}\OperatorTok{.}\AttributeTok{order}\OperatorTok{,}\NormalTok{ portIdentifier}\OperatorTok{,}
\NormalTok{                          channelIdentifier}\OperatorTok{,}
\NormalTok{                          [connection}\OperatorTok{.}\AttributeTok{counterpartyConnectionIdentifier}\NormalTok{]}\OperatorTok{,}
\NormalTok{                          counterpartyVersion\}}
    \FunctionTok{abortTransactionUnless}\NormalTok{(connection}\OperatorTok{.}\FunctionTok{verifyChannelState}\NormalTok{(}
\NormalTok{      proofHeight}\OperatorTok{,}
\NormalTok{      proofTry}\OperatorTok{,}
\NormalTok{      channel}\OperatorTok{.}\AttributeTok{counterpartyPortIdentifier}\OperatorTok{,}
\NormalTok{      channel}\OperatorTok{.}\AttributeTok{counterpartyChannelIdentifier}\OperatorTok{,}
\NormalTok{      expected}
\NormalTok{    ))}
\NormalTok{    channel}\OperatorTok{.}\AttributeTok{state} \OperatorTok{=}\NormalTok{ OPEN}
\NormalTok{    channel}\OperatorTok{.}\AttributeTok{version} \OperatorTok{=}\NormalTok{ counterpartyVersion}
\NormalTok{    provableStore}\OperatorTok{.}\FunctionTok{set}\NormalTok{(}\FunctionTok{channelPath}\NormalTok{(portIdentifier}\OperatorTok{,}\NormalTok{ channelIdentifier)}\OperatorTok{,}\NormalTok{ channel)}
\NormalTok{\}}
\end{Highlighting}
\end{Shaded}

\hypertarget{finalising-a-channel}{%
\subsubsection{Finalising a channel}\label{finalising-a-channel}}

\vspace{3mm}

\begin{Shaded}
\begin{Highlighting}[]
\KeywordTok{function} \FunctionTok{chanOpenConfirm}\NormalTok{(}
\NormalTok{  portIdentifier}\OperatorTok{:}\NormalTok{ Identifier}\OperatorTok{,}
\NormalTok{  channelIdentifier}\OperatorTok{:}\NormalTok{ Identifier}\OperatorTok{,}
\NormalTok{  proofAck}\OperatorTok{:}\NormalTok{ CommitmentProof}\OperatorTok{,}
\NormalTok{  proofHeight}\OperatorTok{:}\NormalTok{ uint64) \{}
\NormalTok{    channel }\OperatorTok{=}\NormalTok{ provableStore}\OperatorTok{.}\FunctionTok{get}\NormalTok{(}\FunctionTok{channelPath}\NormalTok{(portIdentifier}\OperatorTok{,}\NormalTok{ channelIdentifier))}
    \FunctionTok{abortTransactionUnless}\NormalTok{(channel }\OperatorTok{!==}\NormalTok{ null)}
    \FunctionTok{abortTransactionUnless}\NormalTok{(channel}\OperatorTok{.}\AttributeTok{state} \OperatorTok{===}\NormalTok{ TRYOPEN)}
    \FunctionTok{abortTransactionUnless}\NormalTok{(}\FunctionTok{authenticateCapability}\NormalTok{(}\FunctionTok{channelCapabilityPath}\NormalTok{(portIdentifier}\OperatorTok{,}\NormalTok{ channelIdentifier)}\OperatorTok{,}\NormalTok{ capability))}
\NormalTok{    connection }\OperatorTok{=}\NormalTok{ provableStore}\OperatorTok{.}\FunctionTok{get}\NormalTok{(}\FunctionTok{connectionPath}\NormalTok{(channel}\OperatorTok{.}\AttributeTok{connectionHops}\NormalTok{[}\DecValTok{0}\NormalTok{]))}
    \FunctionTok{abortTransactionUnless}\NormalTok{(connection }\OperatorTok{!==}\NormalTok{ null)}
    \FunctionTok{abortTransactionUnless}\NormalTok{(connection}\OperatorTok{.}\AttributeTok{state} \OperatorTok{===}\NormalTok{ OPEN)}
\NormalTok{    expected }\OperatorTok{=}\NormalTok{ ChannelEnd\{OPEN}\OperatorTok{,}\NormalTok{ channel}\OperatorTok{.}\AttributeTok{order}\OperatorTok{,}\NormalTok{ portIdentifier}\OperatorTok{,}
\NormalTok{                          channelIdentifier}\OperatorTok{,}
\NormalTok{                          [connection}\OperatorTok{.}\AttributeTok{counterpartyConnectionIdentifier}\NormalTok{]}\OperatorTok{,}
\NormalTok{                          channel}\OperatorTok{.}\AttributeTok{version}\NormalTok{\}}
    \FunctionTok{abortTransactionUnless}\NormalTok{(connection}\OperatorTok{.}\FunctionTok{verifyChannelState}\NormalTok{(}
\NormalTok{      proofHeight}\OperatorTok{,}
\NormalTok{      proofAck}\OperatorTok{,}
\NormalTok{      channel}\OperatorTok{.}\AttributeTok{counterpartyPortIdentifier}\OperatorTok{,}
\NormalTok{      channel}\OperatorTok{.}\AttributeTok{counterpartyChannelIdentifier}\OperatorTok{,}
\NormalTok{      expected}
\NormalTok{    ))}
\NormalTok{    channel}\OperatorTok{.}\AttributeTok{state} \OperatorTok{=}\NormalTok{ OPEN}
\NormalTok{    provableStore}\OperatorTok{.}\FunctionTok{set}\NormalTok{(}\FunctionTok{channelPath}\NormalTok{(portIdentifier}\OperatorTok{,}\NormalTok{ channelIdentifier)}\OperatorTok{,}\NormalTok{ channel)}
\NormalTok{\}}
\end{Highlighting}
\end{Shaded}

\hypertarget{initiating-channel-closure}{%
\subsubsection{Initiating channel
closure}\label{initiating-channel-closure}}

\vspace{3mm}

\begin{Shaded}
\begin{Highlighting}[]
\KeywordTok{function} \FunctionTok{chanCloseInit}\NormalTok{(}
\NormalTok{  portIdentifier}\OperatorTok{:}\NormalTok{ Identifier}\OperatorTok{,}
\NormalTok{  channelIdentifier}\OperatorTok{:}\NormalTok{ Identifier) \{}
    \FunctionTok{abortTransactionUnless}\NormalTok{(}\FunctionTok{authenticateCapability}\NormalTok{(}\FunctionTok{channelCapabilityPath}\NormalTok{(portIdentifier}\OperatorTok{,}\NormalTok{ channelIdentifier)}\OperatorTok{,}\NormalTok{ capability))}
\NormalTok{    channel }\OperatorTok{=}\NormalTok{ provableStore}\OperatorTok{.}\FunctionTok{get}\NormalTok{(}\FunctionTok{channelPath}\NormalTok{(portIdentifier}\OperatorTok{,}\NormalTok{ channelIdentifier))}
    \FunctionTok{abortTransactionUnless}\NormalTok{(channel }\OperatorTok{!==}\NormalTok{ null)}
    \FunctionTok{abortTransactionUnless}\NormalTok{(channel}\OperatorTok{.}\AttributeTok{state} \OperatorTok{!==}\NormalTok{ CLOSED)}
\NormalTok{    connection }\OperatorTok{=}\NormalTok{ provableStore}\OperatorTok{.}\FunctionTok{get}\NormalTok{(}\FunctionTok{connectionPath}\NormalTok{(channel}\OperatorTok{.}\AttributeTok{connectionHops}\NormalTok{[}\DecValTok{0}\NormalTok{]))}
    \FunctionTok{abortTransactionUnless}\NormalTok{(connection }\OperatorTok{!==}\NormalTok{ null)}
    \FunctionTok{abortTransactionUnless}\NormalTok{(connection}\OperatorTok{.}\AttributeTok{state} \OperatorTok{===}\NormalTok{ OPEN)}
\NormalTok{    channel}\OperatorTok{.}\AttributeTok{state} \OperatorTok{=}\NormalTok{ CLOSED}
\NormalTok{    provableStore}\OperatorTok{.}\FunctionTok{set}\NormalTok{(}\FunctionTok{channelPath}\NormalTok{(portIdentifier}\OperatorTok{,}\NormalTok{ channelIdentifier)}\OperatorTok{,}\NormalTok{ channel)}
\NormalTok{\}}
\end{Highlighting}
\end{Shaded}

\hypertarget{confirming-channel-closure}{%
\subsubsection{Confirming channel
closure}\label{confirming-channel-closure}}

\vspace{3mm}

\begin{Shaded}
\begin{Highlighting}[]
\KeywordTok{function} \FunctionTok{chanCloseConfirm}\NormalTok{(}
\NormalTok{  portIdentifier}\OperatorTok{:}\NormalTok{ Identifier}\OperatorTok{,}
\NormalTok{  channelIdentifier}\OperatorTok{:}\NormalTok{ Identifier}\OperatorTok{,}
\NormalTok{  proofInit}\OperatorTok{:}\NormalTok{ CommitmentProof}\OperatorTok{,}
\NormalTok{  proofHeight}\OperatorTok{:}\NormalTok{ uint64) \{}
    \FunctionTok{abortTransactionUnless}\NormalTok{(}\FunctionTok{authenticateCapability}\NormalTok{(}\FunctionTok{channelCapabilityPath}\NormalTok{(portIdentifier}\OperatorTok{,}\NormalTok{ channelIdentifier)}\OperatorTok{,}\NormalTok{ capability))}
\NormalTok{    channel }\OperatorTok{=}\NormalTok{ provableStore}\OperatorTok{.}\FunctionTok{get}\NormalTok{(}\FunctionTok{channelPath}\NormalTok{(portIdentifier}\OperatorTok{,}\NormalTok{ channelIdentifier))}
    \FunctionTok{abortTransactionUnless}\NormalTok{(channel }\OperatorTok{!==}\NormalTok{ null)}
    \FunctionTok{abortTransactionUnless}\NormalTok{(channel}\OperatorTok{.}\AttributeTok{state} \OperatorTok{!==}\NormalTok{ CLOSED)}
\NormalTok{    connection }\OperatorTok{=}\NormalTok{ provableStore}\OperatorTok{.}\FunctionTok{get}\NormalTok{(}\FunctionTok{connectionPath}\NormalTok{(channel}\OperatorTok{.}\AttributeTok{connectionHops}\NormalTok{[}\DecValTok{0}\NormalTok{]))}
    \FunctionTok{abortTransactionUnless}\NormalTok{(connection }\OperatorTok{!==}\NormalTok{ null)}
    \FunctionTok{abortTransactionUnless}\NormalTok{(connection}\OperatorTok{.}\AttributeTok{state} \OperatorTok{===}\NormalTok{ OPEN)}
\NormalTok{    expected }\OperatorTok{=}\NormalTok{ ChannelEnd\{CLOSED}\OperatorTok{,}\NormalTok{ channel}\OperatorTok{.}\AttributeTok{order}\OperatorTok{,}\NormalTok{ portIdentifier}\OperatorTok{,}
\NormalTok{                          channelIdentifier}\OperatorTok{,}
\NormalTok{                          [connection}\OperatorTok{.}\AttributeTok{counterpartyConnectionIdentifier}\NormalTok{]}\OperatorTok{,}
\NormalTok{                          channel}\OperatorTok{.}\AttributeTok{version}\NormalTok{\}}
    \FunctionTok{abortTransactionUnless}\NormalTok{(connection}\OperatorTok{.}\FunctionTok{verifyChannelState}\NormalTok{(}
\NormalTok{      proofHeight}\OperatorTok{,}
\NormalTok{      proofInit}\OperatorTok{,}
\NormalTok{      channel}\OperatorTok{.}\AttributeTok{counterpartyPortIdentifier}\OperatorTok{,}
\NormalTok{      channel}\OperatorTok{.}\AttributeTok{counterpartyChannelIdentifier}\OperatorTok{,}
\NormalTok{      expected}
\NormalTok{    ))}
\NormalTok{    channel}\OperatorTok{.}\AttributeTok{state} \OperatorTok{=}\NormalTok{ CLOSED}
\NormalTok{    provableStore}\OperatorTok{.}\FunctionTok{set}\NormalTok{(}\FunctionTok{channelPath}\NormalTok{(portIdentifier}\OperatorTok{,}\NormalTok{ channelIdentifier)}\OperatorTok{,}\NormalTok{ channel)}
\NormalTok{\}}
\end{Highlighting}
\end{Shaded}

\hypertarget{packet-handling}{%
\subsection{Packet Handling}\label{packet-handling}}

\hypertarget{sending-a-packet}{%
\subsubsection{Sending a packet}\label{sending-a-packet}}

\vspace{3mm}

\begin{Shaded}
\begin{Highlighting}[]
\KeywordTok{function} \FunctionTok{sendPacket}\NormalTok{(packet}\OperatorTok{:}\NormalTok{ Packet) \{}
\NormalTok{    channel }\OperatorTok{=}\NormalTok{ provableStore}\OperatorTok{.}\FunctionTok{get}\NormalTok{(}\FunctionTok{channelPath}\NormalTok{(packet}\OperatorTok{.}\AttributeTok{sourcePort}\OperatorTok{,}\NormalTok{ packet}\OperatorTok{.}\AttributeTok{sourceChannel}\NormalTok{))}
    \FunctionTok{abortTransactionUnless}\NormalTok{(channel }\OperatorTok{!==}\NormalTok{ null)}
    \FunctionTok{abortTransactionUnless}\NormalTok{(channel}\OperatorTok{.}\AttributeTok{state} \OperatorTok{!==}\NormalTok{ CLOSED)}
    \FunctionTok{abortTransactionUnless}\NormalTok{(}\FunctionTok{authenticateCapability}\NormalTok{(}
      \FunctionTok{channelCapabilityPath}\NormalTok{(packet}\OperatorTok{.}\AttributeTok{sourcePort}\OperatorTok{,}\NormalTok{ packet}\OperatorTok{.}\AttributeTok{sourceChannel}\NormalTok{)}\OperatorTok{,}\NormalTok{ capability))}
    \FunctionTok{abortTransactionUnless}\NormalTok{(packet}\OperatorTok{.}\AttributeTok{destPort} \OperatorTok{===}\NormalTok{ channel}\OperatorTok{.}\AttributeTok{counterpartyPortIdentifier}\NormalTok{)}
    \FunctionTok{abortTransactionUnless}\NormalTok{(packet}\OperatorTok{.}\AttributeTok{destChannel} \OperatorTok{===}\NormalTok{ channel}\OperatorTok{.}\AttributeTok{counterpartyChannelIdentifier}\NormalTok{)}
\NormalTok{    connection }\OperatorTok{=}\NormalTok{ provableStore}\OperatorTok{.}\FunctionTok{get}\NormalTok{(}\FunctionTok{connectionPath}\NormalTok{(channel}\OperatorTok{.}\AttributeTok{connectionHops}\NormalTok{[}\DecValTok{0}\NormalTok{]))}
    \FunctionTok{abortTransactionUnless}\NormalTok{(connection }\OperatorTok{!==}\NormalTok{ null)}
\NormalTok{    latestClientHeight }\OperatorTok{=}\NormalTok{ provableStore}\OperatorTok{.}\FunctionTok{get}\NormalTok{(}\FunctionTok{clientPath}\NormalTok{(connection}\OperatorTok{.}\AttributeTok{clientIdentifier}\NormalTok{))}\OperatorTok{.}\FunctionTok{latestClientHeight}\NormalTok{()}
    \FunctionTok{abortTransactionUnless}\NormalTok{(packet}\OperatorTok{.}\AttributeTok{timeoutHeight} \OperatorTok{===} \DecValTok{0} \OperatorTok{||}\NormalTok{ latestClientHeight }\OperatorTok{\textless{}}\NormalTok{ packet}\OperatorTok{.}\AttributeTok{timeoutHeight}\NormalTok{)}
\NormalTok{    nextSequenceSend }\OperatorTok{=}\NormalTok{ provableStore}\OperatorTok{.}\FunctionTok{get}\NormalTok{(}\FunctionTok{nextSequenceSendPath}\NormalTok{(packet}\OperatorTok{.}\AttributeTok{sourcePort}\OperatorTok{,}\NormalTok{ packet}\OperatorTok{.}\AttributeTok{sourceChannel}\NormalTok{))}
    \FunctionTok{abortTransactionUnless}\NormalTok{(packet}\OperatorTok{.}\AttributeTok{sequence} \OperatorTok{===}\NormalTok{ nextSequenceSend)}
\NormalTok{    nextSequenceSend }\OperatorTok{=}\NormalTok{ nextSequenceSend }\OperatorTok{+} \DecValTok{1}
\NormalTok{    provableStore}\OperatorTok{.}\FunctionTok{set}\NormalTok{(}\FunctionTok{nextSequenceSendPath}\NormalTok{(packet}\OperatorTok{.}\AttributeTok{sourcePort}\OperatorTok{,}\NormalTok{ packet}\OperatorTok{.}\AttributeTok{sourceChannel}\NormalTok{)}\OperatorTok{,}\NormalTok{ nextSequenceSend)}
\NormalTok{    provableStore}\OperatorTok{.}\FunctionTok{set}\NormalTok{(}\FunctionTok{packetCommitmentPath}\NormalTok{(packet}\OperatorTok{.}\AttributeTok{sourcePort}\OperatorTok{,}\NormalTok{ packet}\OperatorTok{.}\AttributeTok{sourceChannel}\OperatorTok{,}\NormalTok{ packet}\OperatorTok{.}\AttributeTok{sequence}\NormalTok{)}\OperatorTok{,}
                      \FunctionTok{hash}\NormalTok{(packet}\OperatorTok{.}\AttributeTok{data}\OperatorTok{,}\NormalTok{ packet}\OperatorTok{.}\AttributeTok{timeoutHeight}\OperatorTok{,}\NormalTok{ packet}\OperatorTok{.}\AttributeTok{timeoutTimestamp}\NormalTok{))}
\NormalTok{\}}
\end{Highlighting}
\end{Shaded}

\hypertarget{receiving-a-packet}{%
\subsubsection{Receiving a packet}\label{receiving-a-packet}}

\vspace{3mm}

\begin{Shaded}
\begin{Highlighting}[]
\KeywordTok{function} \FunctionTok{recvPacket}\NormalTok{(}
\NormalTok{  packet}\OperatorTok{:}\NormalTok{ OpaquePacket}\OperatorTok{,}
\NormalTok{  proof}\OperatorTok{:}\NormalTok{ CommitmentProof}\OperatorTok{,}
\NormalTok{  proofHeight}\OperatorTok{:}\NormalTok{ uint64}\OperatorTok{,}
\NormalTok{  acknowledgement}\OperatorTok{:}\NormalTok{ bytes)}\OperatorTok{:}\NormalTok{ Packet \{}
\NormalTok{    channel }\OperatorTok{=}\NormalTok{ provableStore}\OperatorTok{.}\FunctionTok{get}\NormalTok{(}\FunctionTok{channelPath}\NormalTok{(packet}\OperatorTok{.}\AttributeTok{destPort}\OperatorTok{,}\NormalTok{ packet}\OperatorTok{.}\AttributeTok{destChannel}\NormalTok{))}
    \FunctionTok{abortTransactionUnless}\NormalTok{(channel }\OperatorTok{!==}\NormalTok{ null)}
    \FunctionTok{abortTransactionUnless}\NormalTok{(channel}\OperatorTok{.}\AttributeTok{state} \OperatorTok{===}\NormalTok{ OPEN)}
    \FunctionTok{abortTransactionUnless}\NormalTok{(}
      \FunctionTok{authenticateCapability}\NormalTok{(}\FunctionTok{channelCapabilityPath}\NormalTok{(packet}\OperatorTok{.}\AttributeTok{destPort}\OperatorTok{,}\NormalTok{ packet}\OperatorTok{.}\AttributeTok{destChannel}\NormalTok{)}\OperatorTok{,}\NormalTok{ capability))}
    \FunctionTok{abortTransactionUnless}\NormalTok{(packet}\OperatorTok{.}\AttributeTok{sourcePort} \OperatorTok{===}\NormalTok{ channel}\OperatorTok{.}\AttributeTok{counterpartyPortIdentifier}\NormalTok{)}
    \FunctionTok{abortTransactionUnless}\NormalTok{(packet}\OperatorTok{.}\AttributeTok{sourceChannel} \OperatorTok{===}\NormalTok{ channel}\OperatorTok{.}\AttributeTok{counterpartyChannelIdentifier}\NormalTok{)}
    \FunctionTok{abortTransactionUnless}\NormalTok{(provableStore}\OperatorTok{.}\FunctionTok{get}\NormalTok{(}\FunctionTok{packetAcknowledgementPath}\NormalTok{(packet}\OperatorTok{.}\AttributeTok{destPort}\OperatorTok{,}
\NormalTok{      packet}\OperatorTok{.}\AttributeTok{destChannel}\OperatorTok{,}\NormalTok{ packet}\OperatorTok{.}\AttributeTok{sequence}\NormalTok{) }\OperatorTok{===}\NormalTok{ null))}
\NormalTok{    connection }\OperatorTok{=}\NormalTok{ provableStore}\OperatorTok{.}\FunctionTok{get}\NormalTok{(}\FunctionTok{connectionPath}\NormalTok{(channel}\OperatorTok{.}\AttributeTok{connectionHops}\NormalTok{[}\DecValTok{0}\NormalTok{]))}
    \FunctionTok{abortTransactionUnless}\NormalTok{(connection }\OperatorTok{!==}\NormalTok{ null)}
    \FunctionTok{abortTransactionUnless}\NormalTok{(connection}\OperatorTok{.}\AttributeTok{state} \OperatorTok{===}\NormalTok{ OPEN)}
    \FunctionTok{abortTransactionUnless}\NormalTok{(packet}\OperatorTok{.}\AttributeTok{timeoutHeight} \OperatorTok{===} \DecValTok{0} \OperatorTok{||} \FunctionTok{getConsensusHeight}\NormalTok{() }\OperatorTok{\textless{}}\NormalTok{ packet}\OperatorTok{.}\AttributeTok{timeoutHeight}\NormalTok{)}
    \FunctionTok{abortTransactionUnless}\NormalTok{(packet}\OperatorTok{.}\AttributeTok{timeoutTimestamp} \OperatorTok{===} \DecValTok{0} \OperatorTok{||} \FunctionTok{currentTimestamp}\NormalTok{() }\OperatorTok{\textless{}}\NormalTok{ packet}\OperatorTok{.}\AttributeTok{timeoutTimestamp}\NormalTok{)}
    \FunctionTok{abortTransactionUnless}\NormalTok{(connection}\OperatorTok{.}\FunctionTok{verifyPacketData}\NormalTok{(}
\NormalTok{      proofHeight}\OperatorTok{,}
\NormalTok{      proof}\OperatorTok{,}
\NormalTok{      packet}\OperatorTok{.}\AttributeTok{sourcePort}\OperatorTok{,}
\NormalTok{      packet}\OperatorTok{.}\AttributeTok{sourceChannel}\OperatorTok{,}
\NormalTok{      packet}\OperatorTok{.}\AttributeTok{sequence}\OperatorTok{,}
      \FunctionTok{concat}\NormalTok{(packet}\OperatorTok{.}\AttributeTok{data}\OperatorTok{,}\NormalTok{ packet}\OperatorTok{.}\AttributeTok{timeoutHeight}\OperatorTok{,}\NormalTok{ packet}\OperatorTok{.}\AttributeTok{timeoutTimestamp}\NormalTok{)}
\NormalTok{    ))}
    \FunctionTok{if}\NormalTok{ (acknowledgement}\OperatorTok{.}\AttributeTok{length} \OperatorTok{\textgreater{}} \DecValTok{0} \OperatorTok{||}\NormalTok{ channel}\OperatorTok{.}\AttributeTok{order} \OperatorTok{===}\NormalTok{ UNORDERED)}
\NormalTok{      provableStore}\OperatorTok{.}\FunctionTok{set}\NormalTok{(}
        \FunctionTok{packetAcknowledgementPath}\NormalTok{(packet}\OperatorTok{.}\AttributeTok{destPort}\OperatorTok{,}\NormalTok{ packet}\OperatorTok{.}\AttributeTok{destChannel}\OperatorTok{,}\NormalTok{ packet}\OperatorTok{.}\AttributeTok{sequence}\NormalTok{)}\OperatorTok{,}
        \FunctionTok{hash}\NormalTok{(acknowledgement)}
\NormalTok{      )}
    \FunctionTok{if}\NormalTok{ (channel}\OperatorTok{.}\AttributeTok{order} \OperatorTok{===}\NormalTok{ ORDERED) \{}
\NormalTok{      nextSequenceRecv }\OperatorTok{=}\NormalTok{ provableStore}\OperatorTok{.}\FunctionTok{get}\NormalTok{(}\FunctionTok{nextSequenceRecvPath}\NormalTok{(packet}\OperatorTok{.}\AttributeTok{destPort}\OperatorTok{,}\NormalTok{ packet}\OperatorTok{.}\AttributeTok{destChannel}\NormalTok{))}
      \FunctionTok{abortTransactionUnless}\NormalTok{(packet}\OperatorTok{.}\AttributeTok{sequence} \OperatorTok{===}\NormalTok{ nextSequenceRecv)}
\NormalTok{      nextSequenceRecv }\OperatorTok{=}\NormalTok{ nextSequenceRecv }\OperatorTok{+} \DecValTok{1}
\NormalTok{      provableStore}\OperatorTok{.}\FunctionTok{set}\NormalTok{(}\FunctionTok{nextSequenceRecvPath}\NormalTok{(packet}\OperatorTok{.}\AttributeTok{destPort}\OperatorTok{,}\NormalTok{ packet}\OperatorTok{.}\AttributeTok{destChannel}\NormalTok{)}\OperatorTok{,}\NormalTok{ nextSequenceRecv)}
\NormalTok{    \}}
\NormalTok{    return packet}
\NormalTok{\}}
\end{Highlighting}
\end{Shaded}

\hypertarget{acknowledging-a-packet}{%
\subsubsection{Acknowledging a packet}\label{acknowledging-a-packet}}

\vspace{3mm}

\begin{Shaded}
\begin{Highlighting}[]
\KeywordTok{function} \FunctionTok{acknowledgePacket}\NormalTok{(}
\NormalTok{  packet}\OperatorTok{:}\NormalTok{ OpaquePacket}\OperatorTok{,}
\NormalTok{  acknowledgement}\OperatorTok{:}\NormalTok{ bytes}\OperatorTok{,}
\NormalTok{  proof}\OperatorTok{:}\NormalTok{ CommitmentProof}\OperatorTok{,}
\NormalTok{  proofHeight}\OperatorTok{:}\NormalTok{ uint64)}\OperatorTok{:}\NormalTok{ Packet \{}
\NormalTok{    channel }\OperatorTok{=}\NormalTok{ provableStore}\OperatorTok{.}\FunctionTok{get}\NormalTok{(}\FunctionTok{channelPath}\NormalTok{(packet}\OperatorTok{.}\AttributeTok{sourcePort}\OperatorTok{,}\NormalTok{ packet}\OperatorTok{.}\AttributeTok{sourceChannel}\NormalTok{))}
    \FunctionTok{abortTransactionUnless}\NormalTok{(channel }\OperatorTok{!==}\NormalTok{ null)}
    \FunctionTok{abortTransactionUnless}\NormalTok{(channel}\OperatorTok{.}\AttributeTok{state} \OperatorTok{===}\NormalTok{ OPEN)}
    \FunctionTok{abortTransactionUnless}\NormalTok{(}\FunctionTok{authenticateCapability}\NormalTok{(}
      \FunctionTok{channelCapabilityPath}\NormalTok{(packet}\OperatorTok{.}\AttributeTok{sourcePort}\OperatorTok{,}\NormalTok{ packet}\OperatorTok{.}\AttributeTok{sourceChannel}\NormalTok{)}\OperatorTok{,}\NormalTok{ capability))}
    \FunctionTok{abortTransactionUnless}\NormalTok{(packet}\OperatorTok{.}\AttributeTok{destPort} \OperatorTok{===}\NormalTok{ channel}\OperatorTok{.}\AttributeTok{counterpartyPortIdentifier}\NormalTok{)}
    \FunctionTok{abortTransactionUnless}\NormalTok{(packet}\OperatorTok{.}\AttributeTok{destChannel} \OperatorTok{===}\NormalTok{ channel}\OperatorTok{.}\AttributeTok{counterpartyChannelIdentifier}\NormalTok{)}
\NormalTok{    connection }\OperatorTok{=}\NormalTok{ provableStore}\OperatorTok{.}\FunctionTok{get}\NormalTok{(}\FunctionTok{connectionPath}\NormalTok{(channel}\OperatorTok{.}\AttributeTok{connectionHops}\NormalTok{[}\DecValTok{0}\NormalTok{]))}
    \FunctionTok{abortTransactionUnless}\NormalTok{(connection }\OperatorTok{!==}\NormalTok{ null)}
    \FunctionTok{abortTransactionUnless}\NormalTok{(connection}\OperatorTok{.}\AttributeTok{state} \OperatorTok{===}\NormalTok{ OPEN)}
    \FunctionTok{abortTransactionUnless}\NormalTok{(provableStore}\OperatorTok{.}\FunctionTok{get}\NormalTok{(}\FunctionTok{packetCommitmentPath}\NormalTok{(packet}\OperatorTok{.}\AttributeTok{sourcePort}\OperatorTok{,}
\NormalTok{        packet}\OperatorTok{.}\AttributeTok{sourceChannel}\OperatorTok{,}\NormalTok{ packet}\OperatorTok{.}\AttributeTok{sequence}\NormalTok{))}
           \OperatorTok{===} \FunctionTok{hash}\NormalTok{(packet}\OperatorTok{.}\AttributeTok{data}\OperatorTok{,}\NormalTok{ packet}\OperatorTok{.}\AttributeTok{timeoutHeight}\OperatorTok{,}\NormalTok{ packet}\OperatorTok{.}\AttributeTok{timeoutTimestamp}\NormalTok{))}
    \FunctionTok{abortTransactionUnless}\NormalTok{(connection}\OperatorTok{.}\FunctionTok{verifyPacketAcknowledgement}\NormalTok{(}
\NormalTok{      proofHeight}\OperatorTok{,}
\NormalTok{      proof}\OperatorTok{,}
\NormalTok{      packet}\OperatorTok{.}\AttributeTok{destPort}\OperatorTok{,}
\NormalTok{      packet}\OperatorTok{.}\AttributeTok{destChannel}\OperatorTok{,}
\NormalTok{      packet}\OperatorTok{.}\AttributeTok{sequence}\OperatorTok{,}
\NormalTok{      acknowledgement}
\NormalTok{    ))}
    \FunctionTok{if}\NormalTok{ (channel}\OperatorTok{.}\AttributeTok{order} \OperatorTok{===}\NormalTok{ ORDERED) \{}
\NormalTok{      nextSequenceAck }\OperatorTok{=}\NormalTok{ provableStore}\OperatorTok{.}\FunctionTok{get}\NormalTok{(}\FunctionTok{nextSequenceAckPath}\NormalTok{(packet}\OperatorTok{.}\AttributeTok{sourcePort}\OperatorTok{,}\NormalTok{ packet}\OperatorTok{.}\AttributeTok{sourceChannel}\NormalTok{))}
      \FunctionTok{abortTransactionUnless}\NormalTok{(packet}\OperatorTok{.}\AttributeTok{sequence} \OperatorTok{===}\NormalTok{ nextSequenceAck)}
\NormalTok{      nextSequenceAck }\OperatorTok{=}\NormalTok{ nextSequenceAck }\OperatorTok{+} \DecValTok{1}
\NormalTok{      provableStore}\OperatorTok{.}\FunctionTok{set}\NormalTok{(}\FunctionTok{nextSequenceAckPath}\NormalTok{(packet}\OperatorTok{.}\AttributeTok{sourcePort}\OperatorTok{,}\NormalTok{ packet}\OperatorTok{.}\AttributeTok{sourceChannel}\NormalTok{)}\OperatorTok{,}\NormalTok{ nextSequenceAck)}
\NormalTok{    \}}
\NormalTok{    provableStore}\OperatorTok{.}\FunctionTok{delete}\NormalTok{(}\FunctionTok{packetCommitmentPath}\NormalTok{(packet}\OperatorTok{.}\AttributeTok{sourcePort}\OperatorTok{,}\NormalTok{ packet}\OperatorTok{.}\AttributeTok{sourceChannel}\OperatorTok{,}\NormalTok{ packet}\OperatorTok{.}\AttributeTok{sequence}\NormalTok{))}
\NormalTok{    return packet}
\NormalTok{\}}
\end{Highlighting}
\end{Shaded}

\hypertarget{handling-a-timed-out-packet}{%
\subsubsection{Handling a timed-out
packet}\label{handling-a-timed-out-packet}}

\vspace{3mm}

\begin{Shaded}
\begin{Highlighting}[]
\KeywordTok{function} \FunctionTok{timeoutPacket}\NormalTok{(}
\NormalTok{  packet}\OperatorTok{:}\NormalTok{ OpaquePacket}\OperatorTok{,}
\NormalTok{  proof}\OperatorTok{:}\NormalTok{ CommitmentProof}\OperatorTok{,}
\NormalTok{  proofHeight}\OperatorTok{:}\NormalTok{ uint64}\OperatorTok{,}
\NormalTok{  nextSequenceRecv}\OperatorTok{:}\NormalTok{ Maybe}\OperatorTok{\textless{}}\NormalTok{uint64}\OperatorTok{\textgreater{}}\NormalTok{)}\OperatorTok{:}\NormalTok{ Packet \{}
\NormalTok{    channel }\OperatorTok{=}\NormalTok{ provableStore}\OperatorTok{.}\FunctionTok{get}\NormalTok{(}\FunctionTok{channelPath}\NormalTok{(packet}\OperatorTok{.}\AttributeTok{sourcePort}\OperatorTok{,}\NormalTok{ packet}\OperatorTok{.}\AttributeTok{sourceChannel}\NormalTok{))}
    \FunctionTok{abortTransactionUnless}\NormalTok{(channel }\OperatorTok{!==}\NormalTok{ null)}
    \FunctionTok{abortTransactionUnless}\NormalTok{(channel}\OperatorTok{.}\AttributeTok{state} \OperatorTok{===}\NormalTok{ OPEN)}
    \FunctionTok{abortTransactionUnless}\NormalTok{(}\FunctionTok{authenticateCapability}\NormalTok{(}
      \FunctionTok{channelCapabilityPath}\NormalTok{(packet}\OperatorTok{.}\AttributeTok{sourcePort}\OperatorTok{,}\NormalTok{ packet}\OperatorTok{.}\AttributeTok{sourceChannel}\NormalTok{)}\OperatorTok{,}\NormalTok{ capability))}
    \FunctionTok{abortTransactionUnless}\NormalTok{(packet}\OperatorTok{.}\AttributeTok{destChannel} \OperatorTok{===}\NormalTok{ channel}\OperatorTok{.}\AttributeTok{counterpartyChannelIdentifier}\NormalTok{)}
\NormalTok{    connection }\OperatorTok{=}\NormalTok{ provableStore}\OperatorTok{.}\FunctionTok{get}\NormalTok{(}\FunctionTok{connectionPath}\NormalTok{(channel}\OperatorTok{.}\AttributeTok{connectionHops}\NormalTok{[}\DecValTok{0}\NormalTok{]))}
    \FunctionTok{abortTransactionUnless}\NormalTok{(packet}\OperatorTok{.}\AttributeTok{destPort} \OperatorTok{===}\NormalTok{ channel}\OperatorTok{.}\AttributeTok{counterpartyPortIdentifier}\NormalTok{)}
    \FunctionTok{abortTransactionUnless}\NormalTok{(}
\NormalTok{      (packet}\OperatorTok{.}\AttributeTok{timeoutHeight} \OperatorTok{\textgreater{}} \DecValTok{0} \OperatorTok{\&\&}\NormalTok{ proofHeight }\OperatorTok{\textgreater{}=}\NormalTok{ packet}\OperatorTok{.}\AttributeTok{timeoutHeight}\NormalTok{) }\OperatorTok{||}
\NormalTok{      (packet}\OperatorTok{.}\AttributeTok{timeoutTimestamp} \OperatorTok{\textgreater{}} \DecValTok{0} \OperatorTok{\&\&}
\NormalTok{          connection}\OperatorTok{.}\FunctionTok{getTimestampAtHeight}\NormalTok{(proofHeight) }\OperatorTok{\textgreater{}}\NormalTok{ packet}\OperatorTok{.}\AttributeTok{timeoutTimestamp}\NormalTok{))}
    \FunctionTok{abortTransactionUnless}\NormalTok{(provableStore}\OperatorTok{.}\FunctionTok{get}\NormalTok{(}\FunctionTok{packetCommitmentPath}\NormalTok{(packet}\OperatorTok{.}\AttributeTok{sourcePort}\OperatorTok{,}
\NormalTok{        packet}\OperatorTok{.}\AttributeTok{sourceChannel}\OperatorTok{,}\NormalTok{ packet}\OperatorTok{.}\AttributeTok{sequence}\NormalTok{))}
           \OperatorTok{===} \FunctionTok{hash}\NormalTok{(packet}\OperatorTok{.}\AttributeTok{data}\OperatorTok{,}\NormalTok{ packet}\OperatorTok{.}\AttributeTok{timeoutHeight}\OperatorTok{,}\NormalTok{ packet}\OperatorTok{.}\AttributeTok{timeoutTimestamp}\NormalTok{))}
\NormalTok{    if channel}\OperatorTok{.}\AttributeTok{order} \OperatorTok{===}\NormalTok{ ORDERED \{}
      \FunctionTok{abortTransactionUnless}\NormalTok{(nextSequenceRecv }\OperatorTok{\textless{}=}\NormalTok{ packet}\OperatorTok{.}\AttributeTok{sequence}\NormalTok{)}
      \FunctionTok{abortTransactionUnless}\NormalTok{(connection}\OperatorTok{.}\FunctionTok{verifyNextSequenceRecv}\NormalTok{(}
\NormalTok{        proofHeight}\OperatorTok{,}
\NormalTok{        proof}\OperatorTok{,}
\NormalTok{        packet}\OperatorTok{.}\AttributeTok{destPort}\OperatorTok{,}
\NormalTok{        packet}\OperatorTok{.}\AttributeTok{destChannel}\OperatorTok{,}
\NormalTok{        nextSequenceRecv}
\NormalTok{      ))}
\NormalTok{    \} else}
      \FunctionTok{abortTransactionUnless}\NormalTok{(connection}\OperatorTok{.}\FunctionTok{verifyPacketAcknowledgementAbsence}\NormalTok{(}
\NormalTok{        proofHeight}\OperatorTok{,}
\NormalTok{        proof}\OperatorTok{,}
\NormalTok{        packet}\OperatorTok{.}\AttributeTok{destPort}\OperatorTok{,}
\NormalTok{        packet}\OperatorTok{.}\AttributeTok{destChannel}\OperatorTok{,}
\NormalTok{        packet}\OperatorTok{.}\AttributeTok{sequence}
\NormalTok{      ))}
\NormalTok{    provableStore}\OperatorTok{.}\FunctionTok{delete}\NormalTok{(}\FunctionTok{packetCommitmentPath}\NormalTok{(packet}\OperatorTok{.}\AttributeTok{sourcePort}\OperatorTok{,}\NormalTok{ packet}\OperatorTok{.}\AttributeTok{sourceChannel}\OperatorTok{,}\NormalTok{ packet}\OperatorTok{.}\AttributeTok{sequence}\NormalTok{))}
\NormalTok{    if channel}\OperatorTok{.}\AttributeTok{order} \OperatorTok{===}\NormalTok{ ORDERED \{}
\NormalTok{      channel}\OperatorTok{.}\AttributeTok{state} \OperatorTok{=}\NormalTok{ CLOSED}
\NormalTok{      provableStore}\OperatorTok{.}\FunctionTok{set}\NormalTok{(}\FunctionTok{channelPath}\NormalTok{(packet}\OperatorTok{.}\AttributeTok{sourcePort}\OperatorTok{,}\NormalTok{ packet}\OperatorTok{.}\AttributeTok{sourceChannel}\NormalTok{)}\OperatorTok{,}\NormalTok{ channel)}
\NormalTok{    \}}
\NormalTok{    return packet}
\NormalTok{\}}
\end{Highlighting}
\end{Shaded}

\hypertarget{cleaning-up-packet-data}{%
\subsubsection{Cleaning up packet data}\label{cleaning-up-packet-data}}

\vspace{3mm}

\begin{Shaded}
\begin{Highlighting}[]
\KeywordTok{function} \FunctionTok{cleanupPacket}\NormalTok{(}
\NormalTok{  packet}\OperatorTok{:}\NormalTok{ OpaquePacket}\OperatorTok{,}
\NormalTok{  proof}\OperatorTok{:}\NormalTok{ CommitmentProof}\OperatorTok{,}
\NormalTok{  proofHeight}\OperatorTok{:}\NormalTok{ uint64}\OperatorTok{,}
\NormalTok{  nextSequenceRecvOrAcknowledgement}\OperatorTok{:}\NormalTok{ Either}\OperatorTok{\textless{}}\NormalTok{uint64}\OperatorTok{,}\NormalTok{ bytes}\OperatorTok{\textgreater{}}\NormalTok{)}\OperatorTok{:}\NormalTok{ Packet \{}
\NormalTok{    channel }\OperatorTok{=}\NormalTok{ provableStore}\OperatorTok{.}\FunctionTok{get}\NormalTok{(}\FunctionTok{channelPath}\NormalTok{(packet}\OperatorTok{.}\AttributeTok{sourcePort}\OperatorTok{,}\NormalTok{ packet}\OperatorTok{.}\AttributeTok{sourceChannel}\NormalTok{))}
    \FunctionTok{abortTransactionUnless}\NormalTok{(channel }\OperatorTok{!==}\NormalTok{ null)}
    \FunctionTok{abortTransactionUnless}\NormalTok{(channel}\OperatorTok{.}\AttributeTok{state} \OperatorTok{===}\NormalTok{ OPEN)}
    \FunctionTok{abortTransactionUnless}\NormalTok{(}\FunctionTok{authenticateCapability}\NormalTok{(}
      \FunctionTok{channelCapabilityPath}\NormalTok{(packet}\OperatorTok{.}\AttributeTok{sourcePort}\OperatorTok{,}\NormalTok{ packet}\OperatorTok{.}\AttributeTok{sourceChannel}\NormalTok{)}\OperatorTok{,}\NormalTok{ capability))}
    \FunctionTok{abortTransactionUnless}\NormalTok{(packet}\OperatorTok{.}\AttributeTok{destChannel} \OperatorTok{===}\NormalTok{ channel}\OperatorTok{.}\AttributeTok{counterpartyChannelIdentifier}\NormalTok{)}
\NormalTok{    connection }\OperatorTok{=}\NormalTok{ provableStore}\OperatorTok{.}\FunctionTok{get}\NormalTok{(}\FunctionTok{connectionPath}\NormalTok{(channel}\OperatorTok{.}\AttributeTok{connectionHops}\NormalTok{[}\DecValTok{0}\NormalTok{]))}
    \FunctionTok{abortTransactionUnless}\NormalTok{(connection }\OperatorTok{!==}\NormalTok{ null)}
    \FunctionTok{abortTransactionUnless}\NormalTok{(packet}\OperatorTok{.}\AttributeTok{destPort} \OperatorTok{===}\NormalTok{ channel}\OperatorTok{.}\AttributeTok{counterpartyPortIdentifier}\NormalTok{)}
    \FunctionTok{abortTransactionUnless}\NormalTok{(nextSequenceRecv }\OperatorTok{\textgreater{}}\NormalTok{ packet}\OperatorTok{.}\AttributeTok{sequence}\NormalTok{)}
    \FunctionTok{abortTransactionUnless}\NormalTok{(provableStore}\OperatorTok{.}\FunctionTok{get}\NormalTok{(}\FunctionTok{packetCommitmentPath}\NormalTok{(packet}\OperatorTok{.}\AttributeTok{sourcePort}\OperatorTok{,}
\NormalTok{        packet}\OperatorTok{.}\AttributeTok{sourceChannel}\OperatorTok{,}\NormalTok{ packet}\OperatorTok{.}\AttributeTok{sequence}\NormalTok{))}
               \OperatorTok{===} \FunctionTok{hash}\NormalTok{(packet}\OperatorTok{.}\AttributeTok{data}\OperatorTok{,}\NormalTok{ packet}\OperatorTok{.}\AttributeTok{timeoutHeight}\OperatorTok{,}\NormalTok{ packet}\OperatorTok{.}\AttributeTok{timeoutTimestamp}\NormalTok{))}
\NormalTok{    if channel}\OperatorTok{.}\AttributeTok{order} \OperatorTok{===}\NormalTok{ ORDERED}
      \FunctionTok{abortTransactionUnless}\NormalTok{(connection}\OperatorTok{.}\FunctionTok{verifyNextSequenceRecv}\NormalTok{(}
\NormalTok{        proofHeight}\OperatorTok{,}
\NormalTok{        proof}\OperatorTok{,}
\NormalTok{        packet}\OperatorTok{.}\AttributeTok{destPort}\OperatorTok{,}
\NormalTok{        packet}\OperatorTok{.}\AttributeTok{destChannel}\OperatorTok{,}
\NormalTok{        nextSequenceRecvOrAcknowledgement}
\NormalTok{      ))}
\NormalTok{    else}
      \FunctionTok{abortTransactionUnless}\NormalTok{(connection}\OperatorTok{.}\FunctionTok{verifyPacketAcknowledgement}\NormalTok{(}
\NormalTok{        proofHeight}\OperatorTok{,}
\NormalTok{        proof}\OperatorTok{,}
\NormalTok{        packet}\OperatorTok{.}\AttributeTok{destPort}\OperatorTok{,}
\NormalTok{        packet}\OperatorTok{.}\AttributeTok{destChannel}\OperatorTok{,}
\NormalTok{        packet}\OperatorTok{.}\AttributeTok{sequence}\OperatorTok{,}
\NormalTok{        nextSequenceRecvOrAcknowledgement}
\NormalTok{      ))}
\NormalTok{    provableStore}\OperatorTok{.}\FunctionTok{delete}\NormalTok{(}\FunctionTok{packetCommitmentPath}\NormalTok{(packet}\OperatorTok{.}\AttributeTok{sourcePort}\OperatorTok{,}\NormalTok{ packet}\OperatorTok{.}\AttributeTok{sourceChannel}\OperatorTok{,}\NormalTok{ packet}\OperatorTok{.}\AttributeTok{sequence}\NormalTok{))}
\NormalTok{    return packet}
\NormalTok{\}}
\end{Highlighting}
\end{Shaded}

\pagebreak

\twocolumn

\hypertarget{references}{%
\section*{References}\label{references}}
\addcontentsline{toc}{section}{References}

\hypertarget{refs}{}
\begin{cslreferences}
\leavevmode\hypertarget{ref-polkadot_xcmp}{}%
{[}1{]} Alistair Stewart and Fatemeh Shirazi and Leon Groot Bruinderink,
``Web3 foundation research: XCMP.''
\url{https://research.web3.foundation/en/latest/polkadot/XCMP.html},
May-2020.

\leavevmode\hypertarget{ref-ethereum_2_cross_shard}{}%
{[}2{]} E. 2.0 Contributors, ``Ethereum sharding research compendium:
Cross-shard communication.''
\url{https://notes.ethereum.org/@serenity/H1PGqDhpm?type=view\#Cross-shard-communication},
2020.

\leavevmode\hypertarget{ref-hard_problems_sharding_part_two}{}%
{[}3{]} Near Protocol, ``The authoritative guide to blockchain sharding:
Part 2.''
\url{https://medium.com/nearprotocol/unsolved-problems-in-blockchain-sharding-2327d6517f43},
Dec-2018.

\leavevmode\hypertarget{ref-rfc793}{}%
{[}4{]} ``Transmission Control Protocol.'' RFC 793; RFC Editor,
Sep-1981.

\leavevmode\hypertarget{ref-object_capabilities}{}%
{[}5{]} C. Morningstar, ``What are capabilities?''
\url{http://habitatchronicles.com/2017/05/what-are-capabilities/}, 2017.

\leavevmode\hypertarget{ref-coda_protocol}{}%
{[}6{]} I. Meckler and E. Shapiro, ``Coda: Decentralized cryptocurrency
at scale.''
\url{https://cdn.codaprotocol.com/v2/static/coda-whitepaper-05-10-2018-0.pdf},
2018.

\leavevmode\hypertarget{ref-bitcoin}{}%
{[}7{]} S. Nakamoto, ``Bitcoin: A peer-to-peer electronic cash system.''
2009.

\leavevmode\hypertarget{ref-tendermint_consensus_without_mining}{}%
{[}8{]} Jae Kwon, ``Tendermint: Consensus without mining.''
\url{https://tendermint.com/static/docs/tendermint.pdf}, Sep-2014.

\leavevmode\hypertarget{ref-grandpa_consensus}{}%
{[}9{]} Alistair Stewart, ``GRANDPA finality gadget.''
\url{https://github.com/w3f/consensus/blob/master/pdf/grandpa.pdf},
May-2020.

\leavevmode\hypertarget{ref-hotstuff_consensus}{}%
{[}10{]} M. Yin, D. Malkhi, M. K. Reiter, G. G. Gueta, and I. Abraham,
``HotStuff: BFT consensus in the lens of blockchain.''
\url{https://arxiv.org/pdf/1803.05069}, 2018.

\leavevmode\hypertarget{ref-iavl_plus_tree}{}%
{[}11{]} Tendermint, ``IAVL+ tree: A versioned, snapshottable
(immutable) avl+ tree for persistent data.''
\url{https://github.com/tendermint/iavl}, 2020.

\leavevmode\hypertarget{ref-patricia_tree}{}%
{[}12{]} Ethereum, ``Ethereum modified merkle patricia trie
specification.''
\url{https://github.com/ethereum/wiki/wiki/Patricia-Tree}, 2020.

\leavevmode\hypertarget{ref-ibc_cosmos_sdk}{}%
{[}13{]} Cosmos SDK Contributors, ``The cosmos sdk: X/ibc.''
\url{https://github.com/cosmos/cosmos-sdk/tree/master/x/ibc}, May-2020.

\leavevmode\hypertarget{ref-ibc_rust}{}%
{[}14{]} Informal Systems, ``Rust implementation of ibc modules and
relayer.'' \url{https://github.com/informalsystems/ibc-rs}, May-2020.

\leavevmode\hypertarget{ref-relayer_go}{}%
{[}15{]} Iqlusion, ``Server-side ibc relayer.''
\url{https://github.com/iqlusioninc/relayer}, May-2020.

\leavevmode\hypertarget{ref-game_of_zones}{}%
{[}16{]} GoZ Contributors, ``Game of zones.''
\url{https://goz.cosmosnetwork.dev/}, May-2020.

\leavevmode\hypertarget{ref-map_of_zones}{}%
{[}17{]} Bitquasar \& Ztake, ``Map of zones.''
\url{https://mapofzones.com/}, May-2020.

\leavevmode\hypertarget{ref-substrate}{}%
{[}18{]} P. Technologies, ``Substrate: The platform for blockchain
innovators.'' \url{https://github.com/paritytech/substrate}, 2020.

\leavevmode\hypertarget{ref-cosmos_whitepaper}{}%
{[}19{]} Jae Kwon, Ethan Buchman, ``Cosmos: A network of distributed
ledgers.'' \url{https://cosmos.network/cosmos-whitepaper.pdf}, Sep-2016.

\leavevmode\hypertarget{ref-tendermint_latest_gossip}{}%
{[}20{]} Ethan Buchman, Jae Kwon, Zarko Milosevic, ``The latest gossip
on bft consensus.'' \url{https://arxiv.org/pdf/1807.04938}, Nov-2019.

\leavevmode\hypertarget{ref-ethan_frey_ibc_spec}{}%
{[}21{]} Ethan Frey, ``IBC protocol specification v0.3.1.''
\url{https://github.com/cosmos/ics/blob/master/archive/v0_3_1_IBC.pdf},
Nov-2017.
\end{cslreferences}

\end{document}